\documentclass[a4paper, superscriptaddress]{article}

\usepackage{graphicx} 
\usepackage[english]{babel}
\usepackage[utf8]{inputenc}
\usepackage{amsmath}
\usepackage{amssymb}
\usepackage{graphicx}
\usepackage[margin=1in]{geometry} 
\usepackage{caption}
\usepackage{color}
\usepackage{subcaption}
\usepackage{braket}
\usepackage{setspace}
\usepackage{natbib}
\usepackage{authblk}
\usepackage{url}
\usepackage{bm}
\usepackage{hyperref}
\usepackage{ulem}

\setcitestyle{square,aysep={},yysep={;}}

\begin{document}

\title{Photonic Quantum Computing}
\author{Jacquiline Romero and Gerard Milburn}
\affil{Australian Research Council Centre of Excellence for Engineered Quantum Systems and \\School of Mathematics and Physics, University of Queensland, QLD 4072, Australia}
\date{}
\maketitle


\section*{Summary}
Photonic quantum computation refers to quantum computation that uses photons as the physical system for doing the quantum computation. The field is largely divided between discrete-variable (DV) and continuous-variable (CV) photonic quantum computation. In the former,  quantum information is represented by one or more modal properties (e.g. polarisation) that take on distinct values from a finite set.  Quantum information is processed via operations on these modal properties (e.g. waveplates in the case of polarisation), and eventually measured using single-photon detectors. In CV photonic quantum computation, quantum information is represented by properties of the electromagnetic field that take on any value in an interval (e.g. position).  Both CV and DV implementations have been realized experimentally; each has a unique set of challenges that need to be overcome to achieve scalable universal photonic quantum computation.  It is possible to combine both DV and CV in a hybrid CV-DV fashion to overcome the limitations of either approach. 
\\
\\
\noindent\textbf{Keywords:} qubit, qudit, continuous variable, entanglement, squeezing, measurement, measurement-based quantum computation, universal gates, cluster states, fault tolerance, quantum error correction

\section{Introduction}
Quantum computation is a paradigm that takes advantage of the properties of quantum systems to perform computation. Information embodied in quantum systems follows the rules of quantum mechanics. Using quantum computation, specific computational tasks, e.g. finding the prime factors of a number, can be performed more efficiently. The potential for such an advantage has fueled both theory and practical implementations of quantum computation. 

\begin{figure}
\centering
\includegraphics[scale=0.8]{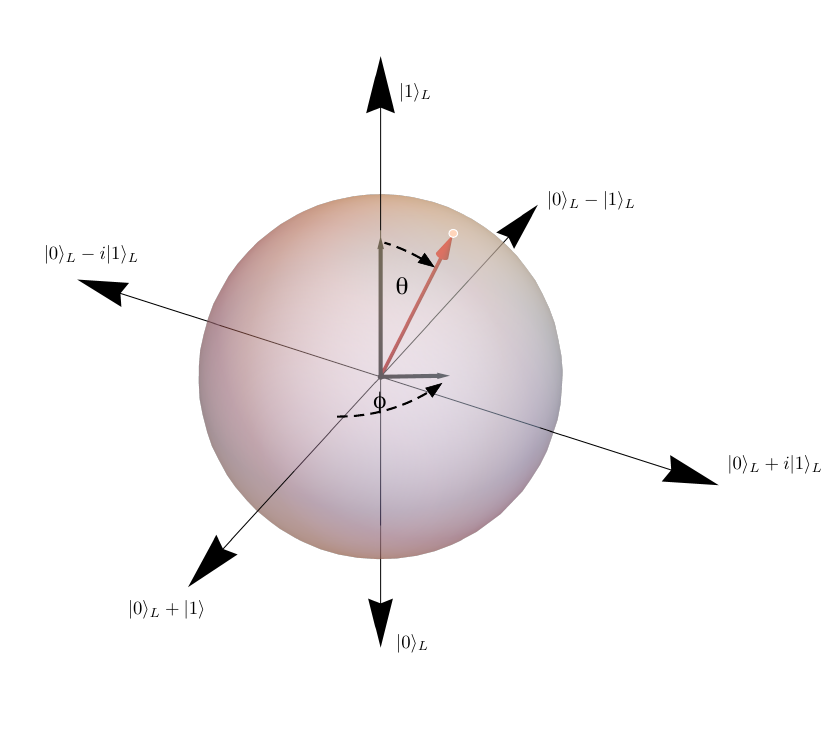}
\caption{A single qubit state is any point on the sphere labelled by two coordinates $\theta,\phi$. This is called the Bloch sphere. If the logical qubits are encoded using two orthogonal polarisation states, this is called the Poincar\'{e} sphere. For example $|0\rangle_L=|H\rangle, |1\rangle_L=|V\rangle$.}
\label{Bloch-sphere}
\end{figure}

Quantum computers operate in a fundamentally different way from conventional computers. Conventional computers are synchronized arrays of cascading irreversible switches (transistors). A stored program sequences the spatio-temporal cascade of conditional switching, and a clock provides the synchronization. The irreversibility originates from the irreversible nature of the binary (on/off) transition in the switch, and the time ordering of the cascade. The switch necessarily dissipates heat.  Quantum computers (as currently conceived)  are synchronised arrays of {\it reversible} gates. A time-order is still imposed (a circuit), but the basic physical units (gates) are reversible (as far as possible) and do not generate heat. However, the computation does not finish until the final output state is measured, and that is necessarily irreversible. If it is not measured, no computation takes place at all. The results of the measurements are binary strings, but the state of the physical devices before measurement are not distinguishable, binary states of matter. They are qubits, and their state is represented by two real numbers ($\phi$ and $\theta$) in one-to-one correspondence with the points on a sphere; see Fig. (\ref{Bloch-sphere}). A point on the sphere represents a superposition state and the numbers $\phi$ and $\theta$ determine the probability of the results of the binary measurement.  During the sequence of gates, the qubits become entangled (highly correlated). Since each qubit is described by two real parameters, the number of real parameters required to describe the state rises exponentially in the number of qubits. It is the ability to control the probability amplitudes (and effectively, the outcome of measurements) 
in this vast space that gives quantum computers their power.   

There are two different kinds of quantum computation (QC) that correspond to whether the qubits are based on matter, or photons. The first kind typically use stationary configurations of atoms, superconducting junctions, trapped ions or spin-polarised degrees of freedom. Photonic QC is based on photonic switching of pulses of light moving along fibres or wave-guides.  Matter QC is much closer to a conventional silicon computer whereas photonic QC is closer to all-optical switching networks as deployed in communication systems, but uses quantum states of light instead of classical. 

There is another fundamental difference between matter and optical QC. The latter lends itself to the protocol of measurement-based QC (MBQC) (\cite{wei2021measurement}) to implement time-ordered, conditional  processing. In MBQC, special highly entangled states of photonic pulses are prepared. Processing is done by measurement and feeding forward the results of the measurement (feed-forward). The reason this is well suited to photonic QC will become more apparent in the  next sections.    

Quantum computers are important because quantum control of matter or photonic qubits using circuits of gates enables new, computationally more efficient, algorithms than are possible using the hardware of a conventional computer. For example,  finding the prime factors of a large integer can be performed exponentially more efficiently. The theory of quantum computation is agnostic to the kinds of physical systems encoding a qubit. They can be matter or photonic. 

The need to keep  quantum circuits as reversible as possible requires quantum error correction. No real physical control of a quantum system is truly reversible since perfect quantum control would require excluding any unknown or unwanted interaction with external degrees of freedom. These ultimately lead to measurement errors (equivalently, noise) at the end of the computation. Were it not for the discovery of powerful methods of quantum error correction, quantum computing would have remained a footnote in the theory of computation. The purpose of error correction is to pump the entropy introduced by unknown or uncontrolled interactions into external degrees of freedom (DOFs) that are ancilliary, hence the name ancila. These ancilas are measured to reveal information about errors, without affecting the logical qubits that are used for the quantum computation. This is necessarily irreversible as the ancila qubits need to first be  prepared in very low entropy states, which costs energy.

 While in the 2020s, noisy intermediate-scale quantum (NISQ) computers that are mostly based on matter systems have been dominant, photonic systems remain a strong contender for a future large-scale universal quantum computer. Whatever physical quantum- computing platform will be ubiquitous in the future, photons will definitely play a major role. Photons are the only quantum information carriers that can ``fly", i.e. they can travel through free space or optical fibres. The compatibility with current optical networks at telecom wavelengths make photons suitable for connecting distributed quantum computers. This article will focus on the development of photonic quantum computing, from early proposals of linear optical quantum computation to the leading contemporary architectures for photonic quantum computing.

\section{Photonic Quantum Information}
\subsection{What is a photon?}
The concept of a photon was introduced by Einstein to explain the photoelectric effect and initiated a decades-long development of quantum field theory.  In the 1960s Glauber (\cite{PhysRev.140.B676,PhysRev.145.1041}),  using the newly developed theory of quantum electrodynamics,  explained how the statistics of photo-electron detection events could be used to understand different quantum states of light. For both Einstein and Glauber the concept of a photon was directly tied to a particular kind of measurement in which discrete detection events --- clicks -- correspond to the absorption of discrete units of electromagnetic energy. As always in quantum mechanics, the picture we draw of the phenomenon under investigation depends on the measurement context. 

Entanglement refers to correlations that do not admit an explanation based on classical physics. Quantum physics allows for entangled quantum states, which upon measurement, lead to correlations that are stronger than that admissible in a classical theory. The local hidden-variable theory that the Bell inequality (\cite{bell1964einstein}) tests against has been shown to be violated in many photonic experiments, most remarkably in the detection-loophole-free experiments of (\cite{shalm2015strong}) and (\cite{giustina2015significant}). The fact that photons interact very weakly with the environment leads to long decoherence times, hence photons are ideal for experiments probing the foundations of quantum physics. Outside fundamental quantum physics, photons can also be the basis of various quantum technologies, e.g. quantum computing.

\subsection{Optical modes}
\label{multimode-states}
What is the Hilbert space of a photon? A photon is a single particle excitation of an optical field {\em mode}.  There is potential for confusion here. Normally, a photon is thought of as a single 'particle' in the sense of non-relativistic quantum mechanics, but in quantum field theory the Hilbert space is associated with orthonormal modes of the electromagnetic (EM) field. 
For each orthonormal mode indexed by $k$ annihilation and creation operators $a_k, a_k^{\dagger}$ where $[a_k, a_k'^{\dagger}]=\delta_{kk'}$ are defined. The electric field can then be written as (\cite{Wal-Mil2008}), 

\begin{equation}
\mathbf{E}(r, t)=i \sum_k\left(\frac{\hbar \omega_k}{2 \epsilon_0}\right)^{1 / 2}\left[a_k \mathbf{u}_k(\mathbf{r}) e^{-i \omega_k t}-a_k^{\dagger} \mathbf{u}_k^*(\mathbf{r}) e^{i \omega_k t}\right]
\end{equation}

\noindent where $\omega_k=c|\vec{k}|$ and $\{\mathbf{u}_k(\mathbf{r})\}$ are spatial mode functions that satisfy

\begin{equation}
\begin{aligned}
\left(\nabla^2+\frac{\omega_k}{c^2}\right) \mathbf{u}_k(\mathbf{r})=0 \\
\int_V \mathbf{u}_k^*(\mathbf{r}) \mathbf{u}_l(\mathbf{r}) d \mathbf{r}=\delta_{k l}.
\end{aligned}
\end{equation}

\noindent The operator with  a discrete spectrum is the number operator for mode $k$, $a_k^{\dagger} a_k$
\begin{equation}
    a^\dagger_k a_k|n\rangle_k =n_k|n\rangle_k\ \ \ n =0,1,2,\ldots
\end{equation}
The operators with a continuous spectrum for mode $k$ are 
\begin{equation}
\label{canonical}
    q_{k}  =\frac{1}{\sqrt{2}}(a_k+a_k^\dagger)\ ,\ \ \ \ \ \  p_{k} = -i\frac{1}{\sqrt{2}}(a_k-a_k^\dagger)
\end{equation}
where $[q_{k}, p_{k'}] = i\delta_{kk'}$.

A single photon state can then be a single excitation of a  single mode or a superposition of a single-photon state of many modes, for example $ |1\rangle =\sum_k f_k a_k^\dagger|0\rangle$ (where $f_k$ are complex coefficients and $a_k^\dagger$ is a creation operator). The Hilbert space is associated with the mode itself.  The Hilbert space of each mode is mathematically equivalent to the Hilbert space of a simple harmonic oscillator. The total Hilbert space of the EM field is the tensor product Hilbert space of many modes.


This can be made more precise by defining a single photon state as an elementary excitation of a spacetime/polarisation mode of the electromagnetic field. In terms of plane waves, each field mode is characterised by a wave vector ${\bm k}$ and a two-dimensional polarisation vector ${\bm \epsilon}$ perpendicular to the wave vector. The energy of a single  excitation (a single photon)  of this mode is given by $\hbar \omega$ where $\omega=c|{\bm k}|$. In the case of a quantum field, the single excitation does not need to be confined to a single mode of the field. A single photon pulse of light will lead to a single detection event distributed over a time interval determined by the temporal shape of the pulse. The photon is delocalised, with a spatial distribution that is proportional to the intensity of the mode  (\cite{loudon2000quantum}).

In quantum optics, the field modes of interest are generally concentrated in a frequency band around an optical carrier frequency that is much larger than the bandwidth. In that case a $1$-photon excitation of a pulse can be approximated in terms of a single-photon excitation superposed over many field modes with wave vector ${\bm k}$ and polarisation ${\bm \epsilon }$. Assuming that only a single spatial direction and polarisation are excited,  this reduces to a frequency integral alone, 
\begin{equation}
\label{number-state}
|\xi\rangle=\int_{-\infty}^\infty d\omega\  \tilde{\xi}(\omega)\tilde{a}^\dagger(\omega)|0\rangle\ \ \ .
\end{equation}
Here $\tilde{\xi}(\omega)$ is the probability amplitude that there is a single photon in a frequency band between $\omega$ and $\omega+d\omega$. Normalisation of the state requires that
\begin{equation}
\int_{-\infty}^\infty d\omega  |\tilde{\xi}(\omega)|^2=1
\end{equation}

 The quantum coherent nature of a single photon is revealed when considering the probability per unit time to detect the photon on a photon counter (the intensity), 
\begin{equation}
\label{intensity}
n(t)=\langle a^\dagger(t)a(t)\rangle =|\xi(t)|^2\ \ \ ,
\end{equation}
where 
\begin{equation}
\xi(t)=\int_{-\infty}^\infty d\omega\ e^{-i\omega t}\tilde{\xi}(\omega )\ \ \\ ;  a(t) =\int_{-\infty}^\infty d\omega \tilde{a}(\omega) e^{-i\omega t}.
\end{equation}
The probability per  unit time to detect a photon depends on the squared modulus of a single, complex-valued function in Eq. (\ref{intensity}).  In optical terms, the pulse is `transform limited' although it should be borne in mind that this is highly non-classical state with an average field amplitude of zero. 

Measurements that result in the number of photons detected (i.e. photon-counting measurements) are not essential. The electric field of a single photon can easily be measured using homodyne detection. This is a phase-dependent measurement of the field and results in a homodyne current $J(t)$:
\begin{equation}
    J(t) =\sqrt{\kappa} \langle a(t) e^{i\theta} + a^\dagger(t) e^{-i\theta})
\end{equation}
where $\kappa$ is a rate constant that depends on the detector. 
It is easy to see that this is zero for a single photon pulse, indicating the random optical phase of a photon-number eigenstate. 
\begin{figure}
    \centering
    \includegraphics[width=0.7\linewidth]{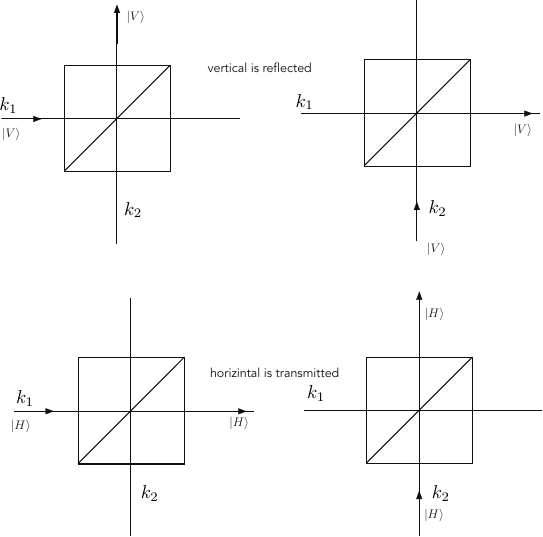}
    \caption{A polarising beamsplitter has two input ports labelled by a wave vector. It reflects vertically-  polarised photons and transmits horizontally-polarised photons.  }
    \label{PBS}
\end{figure}

The definition of a single photon pulse can be extended to include coherent single excitations of different polarisation modes with the same wave vector,
\begin{equation}
\label{number-state}
|\xi\rangle_{k, \epsilon}=\int_{-\infty}^\infty d\omega\  \tilde{\xi}(\omega)\tilde{a}_{k, \epsilon}^\dagger(\omega)|0\rangle\ \ \ .
\end{equation}
where $\epsilon$ labels an arbitrary polarisation direction for each wave vector $k_i$. If this photon is sent into one of the two ports of  a polarising beamsplitter (PBS, see Fig. \ref{PBS}) the field operators are transformed as
\begin{eqnarray}
\tilde{a}_{1,\epsilon}(\omega) & \rightarrow & \sqrt{\eta}\tilde{a}_{1,H}(\omega)+\sqrt{1-\eta}\tilde{a}_{2,V}(\omega)\\
\tilde{a}_{2,\epsilon}(\omega) & \rightarrow & \sqrt{\eta}\tilde{a}_{2,H}(\omega)-\sqrt{1-\eta}\tilde{a}_{1,V}(\omega),
\end{eqnarray}
where all transmitted modes are labelled $H$, all reflected modes are labelled $V$, and $\eta$ is a positive real number less than 1.  The orientation of the PBS defines what is meant by horizontal (H) and vertical (V) in this experiment. If the single photon state $|\xi\rangle_{1,\epsilon}$ is input, the probability per unit time to count a photon at each output channel is
\begin{equation}
    P_1(t) = \eta |\xi(t)|^2 ,\ \ \ \ \  P_2(t) = 1-\eta |\xi(t)|^2
\end{equation}

As a counted photon is absorbed, a detection event can only be obtained at one output, not both.  A convenient notation is to label the state after the PBS as a superposition of a single photon over the two orthogonal polarisations. For example if $|\xi\rangle_{1,\epsilon}$ is input, the output state is written $|\psi\rangle_{out} =\sqrt{\eta}|\xi\rangle_H|0\rangle_V+\sqrt{1-\eta}|0\rangle_H|\xi\rangle_V$. If the same temporal mode is always used, this can be simplified further by omitting the label $\xi$ and all modes in the vacuum and writing $|\xi\rangle_H|0\rangle_V=|H\rangle, |0\rangle |\xi\rangle_V=|V\rangle$ so that 
\begin{equation}
    |\psi\rangle_{out} =\sqrt{\eta}|H\rangle+\sqrt{1-\eta}|V\rangle.
\end{equation}



\begin{table}
    \centering
    \caption{Discrete---qubit ($d{=}2$) and qudit ($d{>}2$)---and continuous photonic quantum information}
    \includegraphics[width=0.8 \columnwidth]{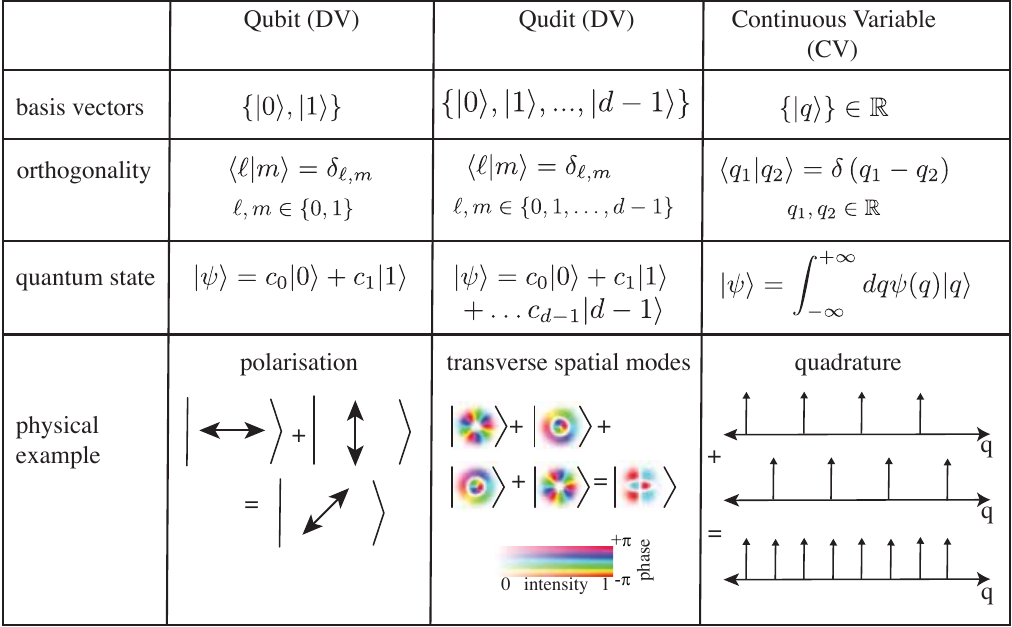}
    \label{qubits}
\end{table}

\subsection{Photonic quantum information: discrete vs continuous}
\label{optical-qubits}
A quantized optical field can be associated with various physical properties. Some of these physical properties are \emph{discrete variables} in that they can take on only distinct values (e.g. the number of photons counted), while some are \emph{continuous variables} (e.g. electric field) and can be any real number. Discrete-variable quantum computation (DVQC) uses discrete variables embodied in discrete properties (as in the number encoding discussed in Sec. \ref{number-encoding} , while continuous-variable quantum computation (CVQC) uses continuous variables embodied in continuous properties (as in the CV-encoding discussed in Sec. \ref{CV-encoding}). DVQC can use either qubits (where the local dimension $d$ is 2), or qudits (where the local dimension is $d{>}2$). Table \ref{qubits} shows the correspondence for two-qubits, qudits, and CV, together with some physical examples using photons.

In DVQC, a quantum state is written as a superposition of orthogonal vectors $\{\ket{j}\}$ that correspond to distinct eigenstates,
\begin{equation}
\ket{\psi}=\sum_j c_j\ket{j}
\end{equation}
where $j$ is an integer that serves as a label for the vectors in the finite set $\{\ket{j}\}$ and $c_j$ are complex coefficients that are normalised, $\sum_i|c_j|^2{=}1$.  In dual-rail encoding, a single qubit is formed of a single photon excitation of any two optical field modes, say the two orthogonal polarisation of a single spatio-temporal mode. All possible (pure) polarisation states can be described as a superposition of this excitation over the two polarisation modes.  The computational basis can be written as $\ket{\mathbf 0}_L=a^\dagger_H\ket{0,0}_{HV}=\ket{1,0}_{HV}=\ket{H}$ and $\ket{\mathbf 1}_L=a^\dagger_V\ket{0,0}_{HV}=\ket{0,1}_{HV}=\ket{V}$. Other photonic properties accommodate a qu$d$it description, i.e. $|\{\ket{j}\}|{=}d$. For example, any (pure) transverse spatial profile can be expressed as a superposition of  a finite number of spatial mode-functions $\{u_j\}$ that are solutions to the wave equation. Although the number of these functions is theoretically infinite, the physical system  (e.g. the physical aperture) imposes a limit on $d$. In a similar way to dual-rail encoding, $d-$rail encoding uses $d$-modes (restricting photon number to 0 or 1) with the computational basis written as 
\begin{eqnarray}
    \ket{\mathbf 0}_L&=&a^\dagger_0\ket{0,...,0,0}_{012...d-1}=\ket{0,...,0,1}_{012...d-1}=\ket{0}\\
    \ket{\mathbf 1}_L&=&a^\dagger_1\ket{0,...,0,0}_{012...d-1}=\ket{0,...,1,0}_{012...d-1}=\ket{1}\\
    ... \nonumber\\
    \ket{\mathbf d-1}_L&=&a^\dagger_{d-1}\ket{0,...,0,0}_{012...d-1}=\ket{1,...,0,0}_{012...d-1}=\ket{d-1}.
\end{eqnarray}
The state of the qudit can then be written simply as $c_0\ket{0}+c_1\ket{1}+...+c_{d-1}\ket{d-1}$. 

In CVQC (\cite{GKP}), a quantum state is written as a superposition over eigenstates of operators with a continuous rather than discrete spectrum, for example position $\hat{q}$ and momentum, $\hat{p}$, of a free particle. These are canonically conjugate operators, so that $[\hat{q},\hat{p}]=i$, where the system of units has been chosen such that the value of Planck's constant is $\hbar=1$. In Dirac notation, the spectral decomposition can be written as  
\begin{equation}
    \hat{q} =\int_{-\infty}^\infty dq\  q|q\rangle\langle q|, \ \ \ \ \     \hat{p} =\int_{-\infty}^\infty dp\  p|p\rangle\langle p|
\end{equation}
where $\langle q'|q\rangle =\delta (q'-q)$,  $\langle p'|p\rangle =\delta (p'-p)$. 
The canonical commutation relations require that 
$
    \langle q|p\rangle =\frac{1}{\sqrt{2\pi}}e^{-iqp}
$.  Given an arbitrary state $|\psi\rangle$ 
\begin{equation}
 \ket{\psi}=\int_{-\infty}^{+\infty}dq \psi(q)=\int_{-\infty}^{+\infty}dp \tilde{\psi}(p) \ket{p}
\end{equation}
where 
\begin{equation}
    \tilde{\psi}(p) =\frac{1}{2\pi} \int_{-\infty}^\infty dq\  e^{-ipq}\psi(q)
\end{equation}
are Fourier-transform pairs. Here, $q,p$ are real variables and $\psi(q), \tilde{\psi}(p)$ are  complex-valued, continuous function of $q,p$ respectively. Note that while the ket $|q\rangle$ is sometimes written here as if it was a state, it is not a physical state. It has delta-function normalisation and does not lie in the Hilbert space of square-integrable functions (functions for which the integral of the square of the absolute value is finite).  This is just a statement of the Heisenberg uncertainty principle: there are no physical states that have zero uncertainty in position or momentum, or more physically, such states cannot be prepared by a finite sequence of physical operations, although zero uncertainty can be approached given sufficient resources. 

\begin{figure}
\centering
\includegraphics[scale=0.6]{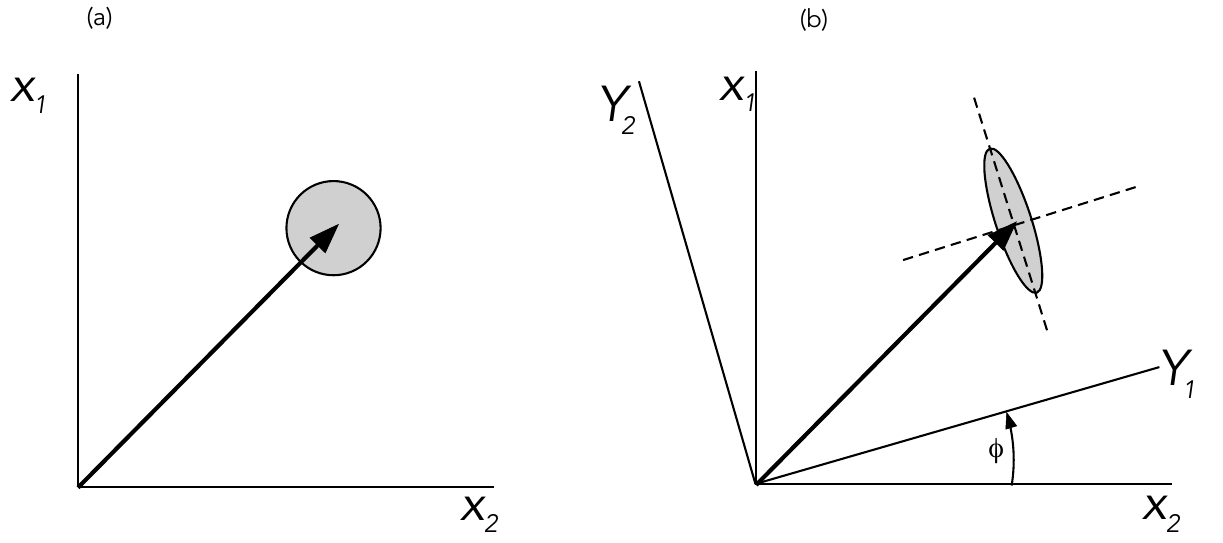}
\caption{A pictorial `phase-space' to capture features of a minimum uncertainty state (a) coherent state (b) squeezed state. }
\label{phase-space-heuristic}
\end{figure}

The squeezed vacuum states are squeezed in that the uncertainty along one quadrature is less than that of the other quadrature, and are defined by
\begin{equation}
|0,\zeta\rangle=S(\zeta)|0\rangle=e^{(\zeta ^*a^2-\zeta a^{\dagger\ 2})/2}\ ,
\end{equation}
where $\zeta= r e^{-2i\phi} $. More generally the squeezed states are defined as displaced squeezed vacuum states
\begin{equation}
|\alpha, \zeta\rangle = D(\alpha)S(\zeta)|0\rangle\ . 
\end{equation}
The moments of the squeezed states can be calculated using the following canonical  transformation 
\begin{equation}\
\label{squeeze-transform}
S^\dagger(\zeta) a S(\zeta) = a\cosh r-a^\dagger e^{-2i\phi}\sinh r
\end{equation}
Defining the rotated quadrature phase operators $Y_1,Y_2$ 
\begin{equation}
Y_1+iY_2 = (X_1+iX_2) e^{-i\phi}\ ,
\end{equation}
and then showing that
\begin{equation}
S^\dagger(\xi) (Y_1+iY_2)  S(\xi) = Y_1 e^{-r} +i Y_2 e^r\ .
\end{equation}
gives a phase-space picture for the state shown in Fig. (\ref{phase-space-heuristic}). 

Light coming from a laser cavity is not in a squeezed state. Squeezing is achieved by pumping a material which has a nonzero second-order optical nonlinear susceptibility (i.e. a $\chi^{(2)}$ nonlinearity). Both the Kerr effect and four-wave mixing can be used to generate squeezed states. Generation of squeezed states relies on degenerate parametric amplifiers---a phase-sensitive amplifier that amplifies noise in one quadrature and deamplifies noise in the other, depending on the phase of the pump. More details can be found in (\cite{bachor2019guide}).



\subsection{Number state qubit encoding}
\label{number-encoding}
\begin{figure}[htbp]
\centering
\includegraphics[width=12cm]{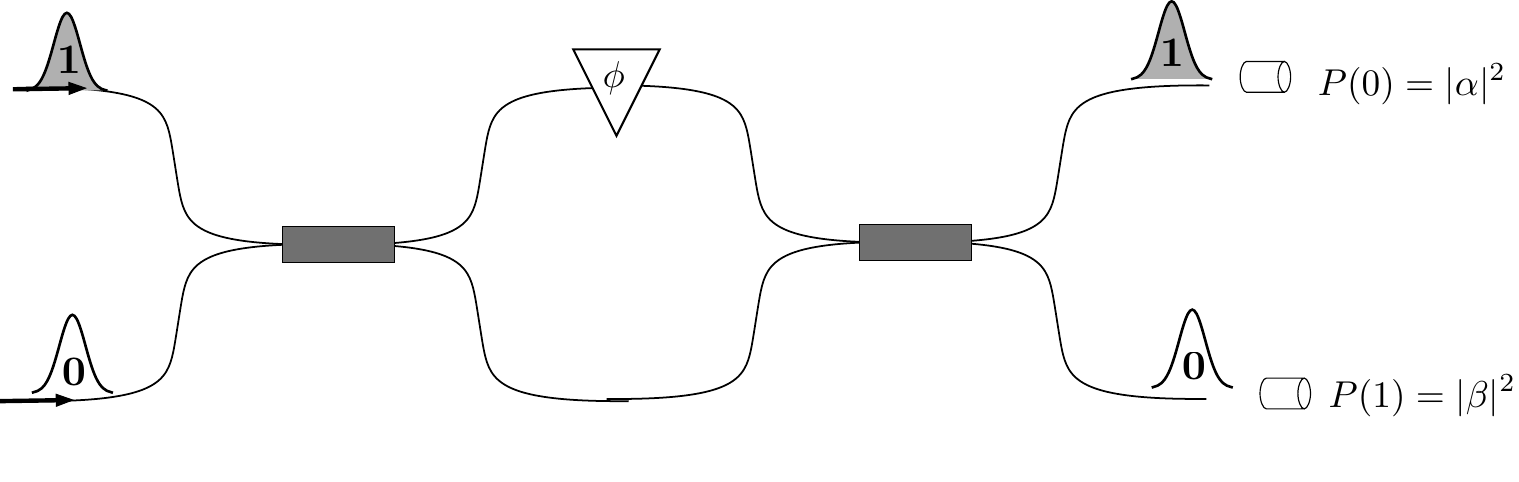}
\caption{An optical realisation of a single qubit gate in an optical fibre (or waveguide)  using dual-rail encoding.  Synchronous transform-limited single-photon pulses can be injected into either.  The relative phase difference is given by $\phi$. Single-photon detectors sample the probability distributions for detection at the upper detector or lower detector. } 
\label{fig_QI8}
\end{figure}

Dual-rail encoding will now be defined more generally. A simple physical qubit based on a single photon pulse of one of a pair of spatio-temporal modes (pulses) will be considered.  Single-photon pulses were described in Sec. (\ref{multimode-states}),
\begin{equation}
|\xi_j\rangle =\int_{-\infty}^\infty d\omega\  \tilde{\xi}(\omega){a}_{j}^\dagger(\omega)|0\rangle
\end{equation}
where $k_j$ labels a particular spatial mode. The explicit reference to $ \tilde{\xi}(\omega)$ will be omitted from now on as it is assumed that all single-photon pulses are prepared in the same temporal mode such that $ |\xi_j\rangle =|1\rangle_j$. The relationship between logical states and the physical photon number state is
\begin{eqnarray}
|{\mathbf 0}\rangle_L & = & |1\rangle_1\otimes|0\rangle_2 =\int_{-\infty}^\infty d\omega\  \tilde{\xi}(\omega){a}_{1}^\dagger(\omega)|0\rangle\\
|{\mathbf 1}\rangle_L & = & |0\rangle_1\otimes|1\rangle_2=\int_{-\infty}^\infty d\omega\  \tilde{\xi}(\omega){a}_{2}^\dagger(\omega)|0\rangle
\end{eqnarray}
The modes could be two input modes to a beam-splitter distinguished by 
the different directions of the wave vector,
or they could be distinguished by polarisation as in the example in Sec. \ref{optical-qubits}. In the case of a beam-splitter, 
 a single qubit gate is easily implemented by the linear transformation
\begin{equation}
a_i(\omega, \theta) =  U(\theta)^\dagger a_i(\omega) U(\theta).
\end{equation}
In order to simplify the notation,  the explicit frequency dependence is dropped from now on but it must be kept in mind that these are defined in frequency space and have the delta-function commutation relations for the free field: 
\begin{equation}
\label{bs-unitary}
  U_{bs}(z)=e^{z a_1 a_2^\dagger -z^* a_1^\dagger a_2}  
\end{equation}
where $z=\theta e^{i\phi}$.
Thus
\begin{eqnarray}
a_1(\theta) & = & \cos\theta a_1+e^{-i\phi}\sin\theta a_2\label{mode_change1} \label{mode_change1}\\
a_2(\theta) & = & \cos\theta a_2-e^{i\phi}\sin\theta a_1\label{mode_change2}.
\label{mode_change2}
\end{eqnarray}
In the case of photons in an optical fibre (or waveguide) as in Fig. \ref{fig_QI8}, $a^\dagger _i$ creates a single-photon excitation at a frequency $\omega$. 
The description in the logical basis becomes,
\begin{eqnarray}
|0\rangle_L& \rightarrow &  \cos\theta_1|0\rangle_L+e^{i\phi}\sin\theta_1|1\rangle_L\\
|1\rangle_L& \rightarrow &  \cos\theta_1|1\rangle_L-e^{-i\phi}\sin\theta_1|0\rangle_L
\end{eqnarray}

A generic qubit state is the  superposition 
\begin{equation}
|\psi\rangle =\alpha |0\rangle_L+ \beta|1\rangle_L
\end{equation}
with $\alpha=\cos\theta, \beta= e^{-i\phi}\sin\theta$. 
The probability to find the system in  one or the other logical state is
\begin{equation}
P({\bf 0})=|\alpha|^2\ , \ \ \ \  P({\bf 1}) =  |\beta|^2
\end{equation}
As these are distinguishable states,  $|\alpha|^2+|\beta|^2=1$.

As $\alpha,\beta\in{\mathbb C}$ (i.e. $\alpha$ and $\beta$ belong to the set of complex numbers $\mathbb C$ ), they are specified by four real numbers. However, normalisation removes one, and another one can be removed by considering only the relative phase $\phi$ between $\alpha$ and $\beta$. This leaves two real numbers to specify the state of a single qubit. Spherical polar coordinates provide a convenient parametrization, 
$\alpha=\cos\theta, \beta =e^{-i\phi}\sin\theta$. A single qubit state is thus in one to one correspondence with  a  point on the  two-sphere, called the Bloch sphere,  as in Fig(\ref{Bloch-sphere}).	This is an infinite continuum of states and an exactly specified state (infinite precision) would require an infinite amount of information. However, despite the state being specified by an infinite amount of information, when the qubit is measured in any basis, only one bit of information is recorded.

\subsection{Continuous-variable qubit encoding }
\label{CV-encoding}
 The first CV encoding scheme in the context of encoding a qubit in the states of a harmonic oscillator,  was proposed by Gottesmann, Kitaev and Preskill (\cite{GKP}). It is known as the GKP encoding. As each optical field mode is equivalent to a quantised harmonic oscillator, this scheme can be used for CV encoding of optical qubits.  A different encoding based on coherent oscillator states was proposed by Ralph et al. (\cite{Ralph2003}). The GKP states are treated first.

\begin{figure}
 \centering
 \includegraphics[scale=0.5]{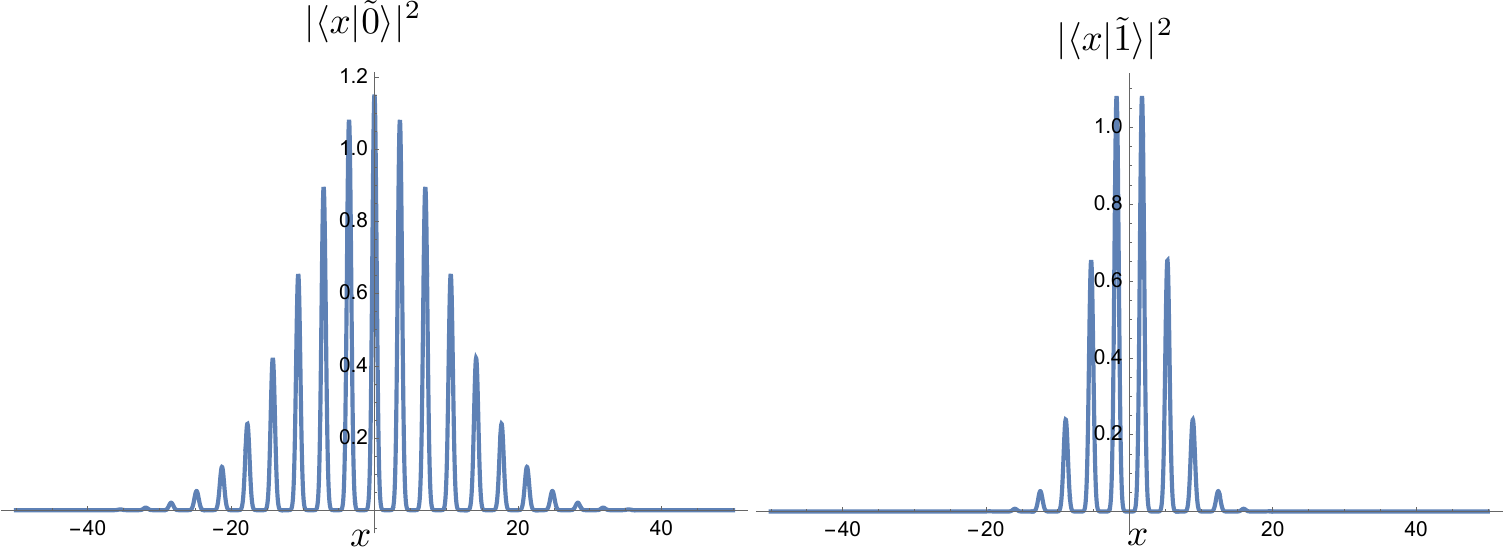}
 \caption{The probability distribution for  $|\tilde{0}\rangle$ (left) and $|\tilde{1}\rangle$ (right ) in the diagonal representation of $X_1=a+a^\dagger$  with $r=1.2, \ \tilde{\Delta}=0.01$ and $|\psi_0\rangle$ has been chosen to be a squeezed state with variance in $X_1$ given by $e^{2r}$.}
 \label{GKPencoding}
 \end{figure}

 In terms of the continuous spectrum of the quadrature phase operator $q$ defined in Eq. (\ref{canonical}), the logical qubits are encoded as (\cite{GKP})

 \begin{equation}
     \begin{array}{l}
|\tilde{0}\rangle=N_0 \sum_{s=-\infty}^{\infty} e^{-\frac{1}{2} \Delta^2(2 s \alpha)^2} T(2 s \alpha)\left|\psi_0\right\rangle, \\
|\tilde{1}\rangle=N_1 \sum_{s=-\infty}^{\infty} e^{\left.-\frac{1}{2}\Delta^2[(2 s+1) \alpha)\right]^2} T[(2 s+1) \alpha]\left|\psi_0\right\rangle,
\end{array}
\label{GKP}
 \end{equation}
\noindent where $N_{0,1}$ are normalization factors, $T(a)$ translates $q$ by $a$, $\alpha$ is a real number, and $\Delta^2$ is the width of the Gaussian envelope that weights the sum of the Gaussians in Eq. \ref{GKP}.  The encoding should ideally consist of infinitely squeezed states in both $q$ and $p$, but this is not physically possible. Instead, the ideal encoding is approximated by a coherent superposition of Gaussian modes as in Eq. \ref{GKP} (the notation $\ket{\tilde{0}}$ and $\ket{\tilde{1}}$ denotes this approximation, Figure \ref{GKPencoding}).  The GKP encoding is of great interest because it is an example of a shift-resistant code, i.e.,  using GKP encoding for an $n-$dimensional system, errors resulting from Pauli operators $\bar{Z}=e^{2 \pi i q / n \alpha}$ and $\bar{X}=e^{-i p \alpha}$ (\cite{GKP}), which cause a shift of $2\pi/n\alpha$ (for $p$) and $\alpha$ (for $q$) respectively, can be protected against shifts.   $\Delta q$ and $\Delta p$ (\cite{GKP}, \cite{glancy2006error}) are given by:
\begin{equation}
    \begin{array}{l}
|\Delta q|<\frac{\alpha}{2}, \\
|\Delta p|<\frac{\pi}{n \alpha} .
\end{array}
\end{equation}

The shift error causes the eigenvalue of the displacement operator to acquire a phase shift, which can be measured and corrected.  This process requires an ancilla qubit and the GKP qubit to be brought to a beamsplitter.  The first output mode undergoes squeezing, while the second output mode is measured.  The result of the measurement is then fed-forward to a shift operation that acts on the first output mode to correct the shift error. The correction prodcedure is described in more detail in (\cite{glancy2006error}).
Konno et al (\cite{Konno}) gave the first experimental demonstration of error correction afforded by GKP states in a travelling-wave optical system using this kind of conditional homodyne measurement and feed-forward. Their scheme only uses linear optics and thus a non-Gaussian resource  state (i.e. a state for which $\psi(q)$ is non-Gaussian) is required. 

Another CV code uses particular superpositions of coherent states called cat states(\cite{Ralph2003}). A laser outputs a coherent state $\ket{\alpha}$, where $\alpha$ is a complex number that determines the average field amplitude. The coherent state $\ket{\alpha}$ arises from a unitary transformation of the vacuum state $\ket{0}$, $\ket{\alpha}=D(\alpha)\ket{0}$,
where $D(\alpha)$ is a displacement operator (in practice, implemented via asymmetric beamsplitters (\cite{paris1996displacement})). These states are not orthogonal; the overlap is given by $|\langle\alpha \mid-\alpha\rangle|^2=e^{-4 \alpha^2}$, but can be very small (e.g. for $\alpha\leq 2$ the overlaps is $\sim 10^{-7}$).  To make an orthogonal encoding, the parity eigenstates can be used:
\begin{equation}
\begin{array}{l}
    \ket{0}=\mathcal{N}_{+}^{-1}\left(\ket{\alpha}+\ket{-\alpha}\right) \\
   \ket{1}=\mathcal{N}_{-}^{-1}\left(\ket{\alpha}-\ket{-\alpha}\right).
\end{array}
\end{equation}
where $\mathcal{N}_{\pm}=2 \pm 2 e^{-2|\alpha|^2}$ \cite{Ralph2003}.  These states are referred to as the even and odd cat states as they only contain even and odd numbers of photons, respectively.  There have been plenty of experimental demonstrations of the generation of cat states, starting with (\cite{ourjoumtsev2006generating}). Cat states were used in the experimental demonstration of GKP states \cite{Konno}.

\section{Photonic Quantum Computing}
\subsection{Preliminaries}
\label{prelim}
This section presents some necessary elements for understanding the next sections. The Hong-Ou-Mandel (HOM) interference is discussed first---a signature quantum effect that cannot be described by classical electromagnetic theory.

HOM interference is a fourth-order interference effect. If two identical single photons arrive on a  $50/50$ beam-splitter, the probability to detect a single photon at each of the two output ports  - a coincidence - is zero.  This is because there are two indistinguishable ways this event can occur: both photons are reflected or both photons are transmitted and the probability amplitudes for each of these paths cancel exactly in the ideal case.  If a time delay between the two photon pulses, the probability to detect a coincidence drops from a value of $0.5$ to zero as the time delay is reduced to zero. 

\begin{figure}[htbp] 
   \centering
   \includegraphics[scale=0.7]{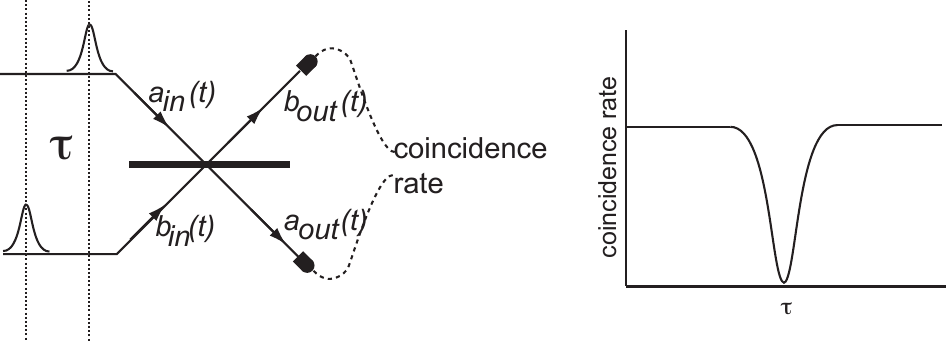} 
   \caption{ A scheme for a Hong-Ou-Mandel interference experiment. A beamsplitter couples two input field modes to two output field modes.  The input field modes are prepared in identical single-photon pulses and a variable time delay $\tau$ is introduced (left).  Detectors placed in the path of the output field modes show a suppression of coincidence events when the delay time  is zero (right). }
   \label{HOM}
\end{figure}

Figure \ref{HOM}, shows a HOM scheme  with two input fields with positive frequency components $a_{in}(t), b_{in}(t)$ and two output fields determined by
\begin{eqnarray}
a_{out}(t)  & = & \frac{1}{\sqrt{2}}( a_{in}(t)+ b_{in}(t))\\
b_{out}(t)  & = & \frac{1}{\sqrt{2}}( a_{in}(t) -b_{in}(t))\ \ \ .
\end{eqnarray}
The change in sign of the second equation is an indication of the time-reversal invariance of an ideal beamsplitter. Assuming that, at $t=0$, the a-mode is prepared in the single-photon state with amplitude function $\xi_a(t)$ while the b-mode is prepared in a single-photon state at $t=\tau$ with amplitude function $\xi_b(t-\tau)$, e.g.
\begin{equation}
\label{sps-function}
\xi(t) = \left \{\begin{array}{ll}  \sqrt{\gamma}e^{-\gamma t/2} & t\geq 0\\
                                                         0 & t< 0
                                                        
                                                         \end{array}
                                                          \right .\ \ \ .
   \end{equation}
The joint  probability to count one photon in output mode-a and one photon in output mode-b is defined by  
\begin{equation}
P_{ab}=\int_0^\infty \int_0^\infty \langle a_{{\rm out}}^\dagger (t)b_{{\rm out}}^\dagger (t')b_{{\rm out}}(t')a_{{\rm out}}(t)\rangle dtdt'\ \ \ .
\label{G2}
\end{equation}  
This is a function of $\tau$ and is given by, 
\begin{equation}
P_{ab}(\tau)=\frac{1}{2}\left (1-e^{-\gamma|\tau|}\right )\ \ \ ,
\end{equation}
giving rise to a dip in the counts as in Fig. \ref{HOM}.
HOM interference is important for the entangling gates that will be discussed in Sec. \ref{gates}. 

The single photons used to demonstrate HOM interference often come from spontaneous parametric down-conversion (SPDC), a nonlinear process in which a pump photon is probabilistically transformed into two lower-energy photons (\cite{Boyd}). SPDC is a convenient source of very pure entangled photons, as used for the detection loophole-free violations of the Bell inequality (\cite{giustina2015significant}, \cite{shalm2015strong}). Spontaneous four-wave mixing (SFWM) is another  nonlinear process used for generating entangled photons (\cite{Boyd}), often implemented on-chip where the nonlinearities of materials like silicon and silicon nitride can be exploited (\cite{alexander2024manufacturable}). In SFWM, two pump photons are converted to two entangled photons. 

For CV optical quantum computation, the generation of quadrature-squeezed states is usually achieved via optical parametric oscillators. Here, the parametric down-conversion process occurs inside a cavity, thus enhancing the effective nonlinearity. The level of squeezing has steadily increased (from $~0.3$ dB  to $~15$ dB) (\cite{andersen201630}), by significant technical improvements of low-noise electronics, low-loss optical components, and high-efficiency photodetectors  (\cite{bachor2019guide}).

The basic operations on a single qubit are expressed in terms of Pauli matrices: bit flip ($\sigma_x$), phase flip ($\sigma_z$), and a combination of the two ($\sigma_y$):

\begin{equation}
\label{pauli}
\sigma_x  =\left(\begin{array}{ll}
0 & 1 \\
1 & 0
\end{array}\right) ;
\sigma_y  =\left(\begin{array}{cc}
0 & -i \\
i & 0
\end{array}\right);
\sigma_z  =\left(\begin{array}{cc}
1 & 0 \\
0 & -1
\end{array}\right)
\end{equation}
For polarisation, these operations are implemented using half- and quarter- waveplates. For example, a bit flip for horizontal polarisation can be achieved by a half-waveplate oriented at $45^\circ$.  

\subsection{Optical gates}
\label{gates}
In Sec. \ref{optical-qubits}, a number state encoding for optical qubits was defined. For this encoding to be useful, one should be able to generate such states and be able to perform arbitrary single qubit unitary transformations. The generation of such states is discussed below. Single qubit transformations are relatively easy. Suppose the two modes used in the encoding are two distinguishable  spatio-temporal  modes at the input to an arbitrary  beamsplitter described by the unitary transformation in Eq. (\ref{bs-unitary}). (The phase convention is fixed by setting $\phi=0$). The input ($a_k$) and output ($\bar{a}_k$) modes are related by
\begin{eqnarray}
    \bar{a}_{1} & = & \cos\theta  a_{1}+\sin\theta  a_{2}\\
    \bar{a}_{2} & = & \cos\theta  a_{2}-\sin\theta a_{1}
\end{eqnarray}
where $\eta =\cos^2\theta $ is the intensity transmittivity coefficient for each mode. 
This shows that
\begin{eqnarray}
\label{BS1}
    U_{bs}(\theta)|{\bf 0}\rangle  & = &  \cos\theta |{\bf 0}\rangle  +\sin\theta |{\bf 1}\rangle\\
    \label{BS2}
    U_{bs}(\theta)|{\bf 1}\rangle  & = &  \cos\theta |{\bf 1}\rangle  -\sin\theta |{\bf 0}\rangle 
    \end{eqnarray}
 is a single qubit transformation that will be referred to as the beamsplitter transformation. Any additional phase shifts can be accounted for after the beamsplitter. The unitary operator 
\begin{equation}
\label{phase-shift}
  U_p(\phi) = e^{-i\phi (a_1^\dagger a_1-a_2^\dagger a_2)}  
\end{equation}
  is another single qubit gate defined by 
  \begin{eqnarray}
    U_p(\phi)|{\bf 0}\rangle  & = &  e^{-i\phi} |{\bf 0}\rangle \\
    U_p(\phi)|{\bf 1}\rangle  & = &   e^{i\phi} |{\bf 1}\rangle
\end{eqnarray}
  This is called the phase gate as it introduces a relative phase of $2\phi$ between the two modes. A combination of the beamsplitter and relative- phase gates generate an arbitrary single qubit state. It is easy to see that both these gates leave the total photon number invariant. Such gates are referred to as {\em linear optical gates}.

In order to produce an arbitrary state in the tensor-product Hilbert space of two qubits, a two mode transformation that entangles two dual-rail qubits is required. It was  discovered many years ago that this would require a nonlinear optical process such as four-wave mixing(\cite{Yamamoto, Mil-Fredkin}). Unfortunately, such optical nonlinearities are very small at single-photon intensities.  Two measurement-based  protocols were discovered that overcame this problem.   

Knill, Laflamme and Milburn (KLM) (\cite{KLM}) discovered that conditional states produced by conditioning on single-photon counting (as in Fig. \ref{fig_QI8}) are sufficient to entangle photonic qubits when all other gates are linear optical gates (they conserve total photon number). This led to the first proposal for linear optical quantum computing (LOQC).   Pittman, Jacobs and Franson (\cite{PJF}) discovered a different approach based on conditional photon counting that exploits the ability for spontaneous parametric down conversion (SPDC) to entangle two optical modes.   This does not conserve total photon number but it is well understood and easy to implement. It is also the key idea for fusion gates (\cite{BR}), currently the most practical way to do photonic quantum computing. The LOQC scheme will be described first. 

Linear in LOQC refers to the use of linear optical elements, e.g. beamsplitters and phase gates, which act linearly on the input modes $\hat{a}_k^{\dagger}$.  The output of a linear optical element is given by the sum
\begin{equation}
\hat{b}_j^{\dagger}=\sum_k M_{j k} \hat{a}_k^{\dagger}
\end{equation}
where $M_{jk}$ are elements of a unitary matrix. This does not mix creation and annihilation operators so conserves the total photon number.  Because any single-qubit state can be generated by a combination of a beamsplitter and a phase gate, it follows that this combination can also be used for any single-qubit gate.

\begin{figure}[htbp]
\centering
\includegraphics[scale=0.6]{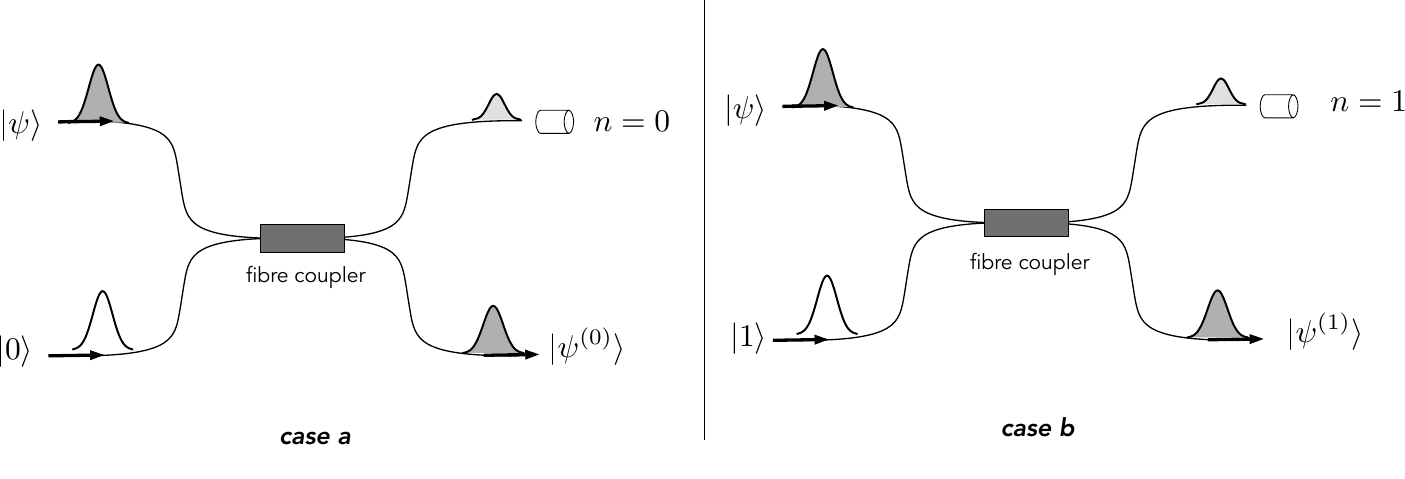}
\caption{A conditional state transformation conditioned on photon- counting measurements. In  case $a$ an optical pulse in an arbitrary quantum state is injected into the top rail while the  bottom rail is a vacuum state. The states input in the bottom rail are either a vacuum state (left) or a single photon state (right).  The conditional state of the bottom rail, given no photons are counted in the top rail is $|\psi^{(0)}\rangle$. In  case $b$ an optical pulse in an arbitrary quantum state is injected into the top rail and a single photon pulse is applied at the  bottom rail. The conditional state of the bottom rail, given one photon is counted in the top rail, is $|\psi^{(1)}\rangle$.   } \label{conditional-state}
\end{figure}

A quite different approach to achieve large single-photon conditional phase shifts is based on the non-unitary transformation of a state that results when a measurement is made. Consider the situation  shown in Fig. \ref{conditional-state}. Two modes of an optical field are coupled via a beamsplitter. One mode is assumed to be in the vacuum state (a) or a one-photon state (b), while the other mode is arbitrary. A single photon counter is placed in the output port of  the top rail. What is the conditional state of the bottom rail given a count of $n$ photons in the top rail? In the example in Fig. \ref{conditional-state}, the state of the bottom rail is $\ket{\psi^{(0)}}$, if there are no photons ($n{=}0$) detected in the top rail. Similarly, the state of the bottom rail is $\ket{\psi^{(1)}}$, if there is a photon ($n{=}1$) detected in the top rail.  Such states are said to be conditioned, because the knowledge of the state in the bottom rails depends on measurement results on the top rail; these states are used in the KLM protocol.

The KLM protocol introduced a way of implementing two-qubit gates using only single-qubit gates, single-photon sources, and detectors that can resolve 0,1, and 2 photons, together with post-selection, a form of measurement-based control. The protocol is run the same way each time, but post-selection is used to flag situations where a gate may not always apply the correct transformation.  The failure is `heralded', that is to say, the actual count implies that an error has occurred. Only the results when a failure was \emph{not} heralded are kept.  This is called post-selection. KLM proposed using post-selection to prepare entangled states as a resource for implementing gates. This is possible using gate-teleportation.

Teleportation is a quantum communication scheme whereby an unknown quantum state can be transmitted from a source to a receiver using only two bits of classical communication (these two bits are used to denote which of the four two-photon maximally entangled states---a Bell pair---have been measured locally on the side of the source; for more details, see \cite{bennett1993teleporting}) and without either sender or receiver learning any information about the transmitted quantum state. In the case of a single qubit, teleportation requires that source and receiver first each share one component of a Bell pair. Making a joint Bell measurement on the qubit state to be transmitted and the source Bell pair component, results in four equally likely outcomes. Sending these two bits to the receiver enables the Bell pair component held by the receiver to be prepared in the unknown quantum state held at the source using a local unitary transformation conditioned by the two bits of classical information. In effect this is a teleportation of the identity gate. By a suitable choice of the shared quantum-entangled state, the unknown quantum state can be transmitted with a single qubit gate applied. This can be extended to the teleportation of entangled states of two or more qubits. In the KLM scheme the required resource state that is shared  is generated nondeterministically by a heralded event.  

Generating heralded entangled states of $n$ photons in $2n$ modes can then be used to make gate teleportation near deterministic at the expense of many optical components and many trials, see (\cite{KLM}) for the details. The basic idea is to have many nondeterministic sources of the required resource state running in parallel; only the one that works is used as a resource to complete the teleportation protocol. This requires fast optical switching.   The resource overhead for gate teleportation is high due to the need for the parallel resource generation.  

The starting point is to nondeterministically implement the nonlinear sign shift (NS) gate which transforms $\alpha_0|0\rangle+\alpha_1|1\rangle+\alpha_2|2\rangle \rightarrow \alpha_0|0\rangle+\alpha_1|1\rangle-\alpha_2|2\rangle$. The $NS$ transformation is nonlinear, but KLM showed that this is possible with just linear optics using two extra modes (the ancilla modes), and photodetection. Dual-rail encoding uses at most one photon per mode for encoding. Implementing the NS gate moves outside the code space but this is easily done using Hong-Ou-Mandel (HOM)(\cite{HOM}) interference. 

\begin{figure}[h!]
    \centering
    \includegraphics[width=0.75 \columnwidth]{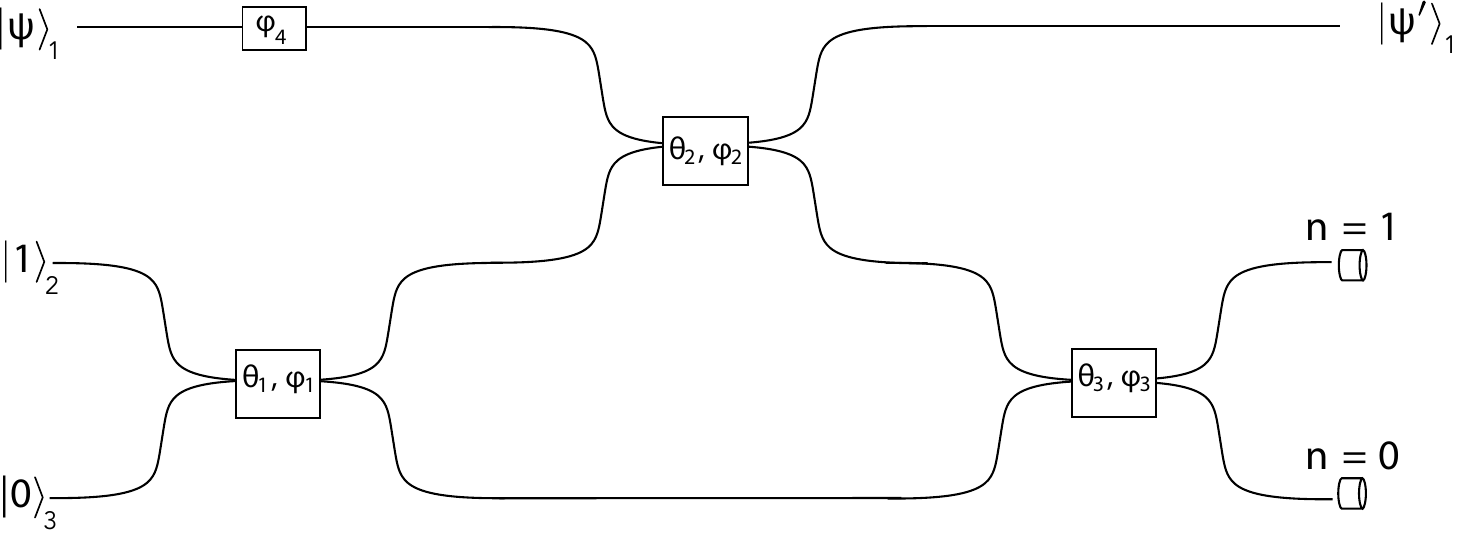}
    \caption{ Nonlinear sign shift gate (\cite{KLM}). The input to mode 1 (top input) is $\ket{\psi}_1=\alpha_0\ket{0}_1+\alpha_1\ket{1}_1+\alpha_2\ket{2}_1$. If a photon is detected in mode 2, and no photon is detected in mode 3, the output of the gate is $NS \ket{\psi}_1=\alpha_0\ket{0}_1+\alpha_1\ket{1}_1-\alpha_2\ket{2}_1$. The success probability of the gate is $1/4$. The values of $\theta_1$, $\theta_2$, $\theta_3$ define the splitting ratio between the paths according to Eqs. \ref{BS1},\ref{BS2}   and are set to $22.5^{\circ}$, $65.5^{\circ}$ ,$22.5^{\circ}$, respectively.  The values of $\phi_1$, $\phi_2$, $\phi_3$ define the relative phase between the paths, and are all set to $0^{\circ}$, while $\phi_4=180^{\circ}$.}
    \label{NS}
\end{figure}

In Fig. \ref{NS}, the top input (mode 1) corresponds to $\ket{\psi}_1=\alpha_0\ket{0}_1+\alpha_1\ket{1}_1+\alpha_2\ket{2}_1$, and modes 2 and 3 correspond to the ancilla modes.  When the input to the NS gate is $\ket{\psi}_1\otimes\ket{1}_2\otimes\ket{0}_3$ (where $\otimes$ denote the tensor product, i.e. for this particular input state, $\ket{\psi}_1, \ket{1}_2, \ket{0}_3$ are fed to input modes 1,2, and 3 respectively, as denoted by the subscripts), and the angles in Fig. \ref{NS}. The single qubit unitary in each box in Fig. \ref{NS} is given by Eqs. \ref{BS1}, \ref{BS2}, and \ref{phase-shift},  with the parameters chosen as $\theta_1=22.5^{\circ}, \phi_1=0^{\circ}, \theta_2=65.5302^{\circ}, \phi_2=0^{\circ}, \theta_3=-22.5^{\circ}, \phi_3=0^{\circ}$ and $\phi_4=180^{\circ}$. The output in mode 1 is $NS\ket{\psi}_1=\alpha_0\ket{0}_1+\alpha_1\ket{1}_1-\alpha_2\ket{2}_1$, as long as the output on mode 2 is $\ket{1}_2$ and the output on mode 3 is $\ket{0}_3$.  The probability of success  with this post-selection is $1/4$.  

\begin{figure}[h!]
    \centering
    \includegraphics[width=0.75 \columnwidth]{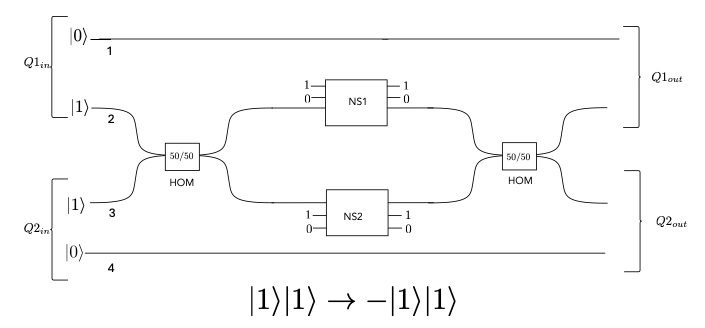}
    \caption{Control-Z  gate (\cite{KLM}). Two NS gates can be used to make a CZ gate by exploiting the HOM effect. The gates acts on $Q_1$ and $Q_2$ (dual-rail qubits, where the modes are labelled as 1 to 4 from top to bottom) as $\ket{Q_1}\ket{Q_2}\rightarrow(-1)^{Q_1 Q_2}\ket{Q_1}\ket{Q_2}$. The first beam-splitter is 50/50 ($\theta=45^\circ$ in Eq. \ref{BS1}, \ref{BS2} ), leads to the state $|02\rangle_{23}+|20\rangle_{23}$ when there is 1 photon input to modes 2 and 3. The second beam-splitter undoes this Hong-Ou-Mandel interference after the action of the two NS gates. The success probability is $1/16$. This is referred to as the  $CZ_{1/16}$. }
    \label{CZ}
\end{figure}

The NS gates are important because two of them make up a non-deterministic conditional sign flip, or a CZ gate: $\ket{Q_1}\ket{Q_2}\rightarrow(-1)^{Q_1 Q_2}\ket{Q_1}\ket{Q_2}$, where $Q_1$ and $Q_2$ are dual-rail-encoded qubits Fig. \ref{CZ}. The CZ gate comprising of two NS gates succeeds $1/16$ of the time, but the times that the gate fails can be determined easily because those are just the times when the ancilla modes of the NS gate are not $\ket{10}$. CZ gates and single-qubit gates can be used to implement CNOT gates, a universal quantum computation gate. 

The KLM scheme pointed the way to quantum computing using photon but requires a huge number of optical elements and control electronics.  For example, a single entangling gate for two qubits that succeeds with probability of $95\%$ requires about 300 successful $CZ_{9/16}$ gate operations (where $9/16$ is the probability of success). The $CZ_{9/16}$ is a generalisation of the scheme in Fig. (\ref{CZ}) that uses a more complicated teleportation resource (\cite{KLM}).  

 At about the same time as KLM, Pittman, Jacobs and Franson (\cite{PJF}) made use of a readily available photonic entangling resource --- spontaneous parametric down conversion (SPDC) --- to implement a different kind of conditional gate.  Nielsen(\cite{PhysRevLett.93.040503}), building on KLM and the concept of measurement-based quantum computation (MBQC)(\cite{MBQC}, \cite{wei2021measurement}), proposed a simpler scheme that avoided gate teleportation.   Browne and Rudolph (\cite{BR}), used SPDC sources together with Nielsen's scheme to give a protocol that dramatically reduced the resources required for optical quantum computing, making it a practical technology for quantum computation.

\begin{figure}
\centering
\includegraphics[scale=0.45]{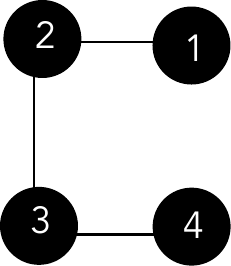}
\caption{A cluster state example. Each logical qubit (dual rail encoding) is represented by a node in the graph. A CZ gate is applied between pairs of physical qubits connected by an edge. Here $
CZ_{12}CZ_{23}CZ_{34} |++++\rangle =\frac{1}{2}\left (|0+0+\rangle+|0-1-\rangle+|1-0+\rangle+|1+1-\rangle\right )$ }
\label{cluster1}
\end{figure}

\begin{figure}
\centering
\includegraphics[scale=0.65]{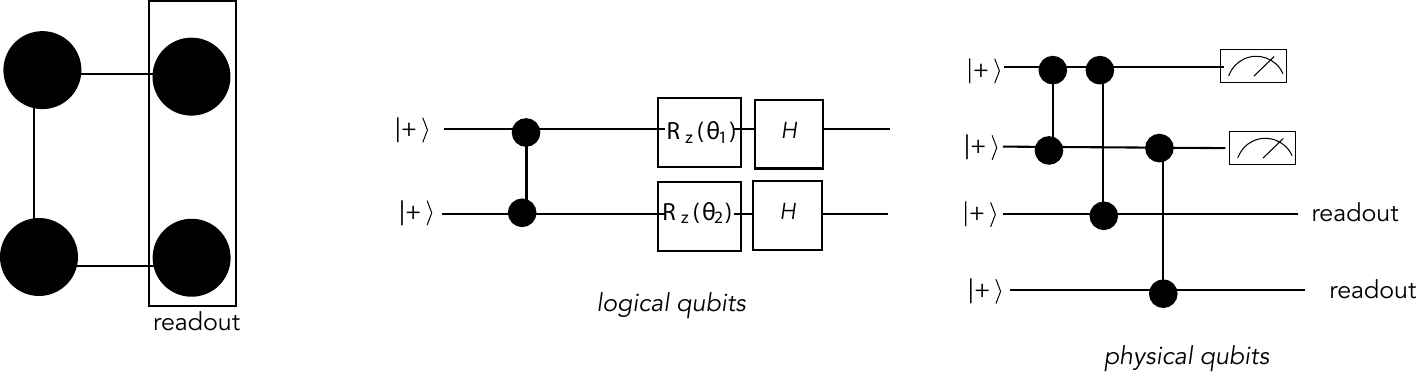}
\caption{The equivalence between a cluster state for MBQC (left) and a quantum circuit for logical qubits (middle). The last vertex in a cluster row is the readout qubit. The right-hand picture shows the creation of the cluster followed by measurements in the computational basis. Depending on the outcome of the measurements, the logical circuit is implemented where the Z-rotation gate parameter (i.e. the Z-rotation angle) depends on the measurement outcomes.  For example if the measurement outcome is $00$, $\theta_1 =\pi, \theta_2=0$  The single qubit rotation gate is defined by  $R_z(\theta)|+\rangle=|0\rangle+e^{i\theta}|1\rangle  $}
\label{cluster3}
\end{figure}

MBQC is implemented on a special class of entangled states called cluster states.  First take $N$ qubits all prepared in the logical $|0\rangle$ state. Represent each dual-rail-encoded qubit as a node in a two dimensional graph (i.e. each qubit is a node and the edges represent entanglement). Apply a H gate to each one, ${\rm H}: |0\rangle\rightarrow |0\rangle+|1\rangle$, and then apply a CZ gate between all qubits connected by an edge. An example is shown in Fig. \ref{cluster1} for a four-qubit cluster state. Each row of a cluster represents one {\em logical qubit}. In the example of Fig. \ref{cluster1}, two physical qubits represent a single logical qubit in each horizontal line, coupled by the vertical edge. 
Cluster-state computation requires making single-qubit measurements on one or more qubits in a cluster and then applying a conditional single-qubit gate on the remaining qubits, conditional on the outcome of the measurements.  For example, given the cluster state 
\begin{equation}
\label{init-cnot}
	|\psi_{\mathrm{in}}\rangle =|i_{1}\rangle_{1} \otimes |+\rangle_{2} \otimes |+\rangle_{3} \otimes |i_{4}\rangle_{4}\ .
\end{equation}
where  $i_{1},i_{4}\in \{0,1\}$, a cluster state can be created by applying the CZ gate,
    \begin{equation}
         G = \exp \left({-i \frac{\pi}{4} \left(\sigma_{z}^{(1)}\sigma_{z}^{(2)} +
        \sigma_{z}^{(2)}\sigma_{z}^{(3)} +
        \sigma_{z}^{(2)}\sigma_{z}^{(4)}\right)}\right).
    \end{equation}
  As the state in Eq. (\ref{init-cnot}) has no $Z$-correlations, when any qubit in the cluster state is measured, it is simply disconnected from the cluster, and does not change the remaining links.     
A $\sigma_x$ measurement is made on qubit 1 and  qubit 2. The measurement outcomes are labelled, $s_j\in\{0,1\}$. The conditional states are
\begin{equation}
	|\psi_{\mathrm{out}}\rangle =|s_{1}\rangle_{1} \otimes |s_{2}\rangle_{2} \otimes U_{\mathrm{con}}|i_{1}+i_{4}\text{ mod } 2\rangle_{3} \otimes |i_{4}\rangle_{4},
\end{equation}
where the conditional single-qubit control
\begin{equation}
 	U_{\mathrm{con}} = \frac{1}{\sqrt{2}}(1+(-1)^{s_{2}}i)\mathrm{e}^{-i\frac{\pi}{4} 		(1+\sigma_{y}^{(3)})}\left( {\sigma_{x}^{(3)}}\right)^{s_{2}}    
\end{equation}
is applied. This implements a  CNOT of the two input qubit states up to a local control, with the output encoded on the third qubit in the cluster.   Given a quantum circuit for the four dual-rail encoded qubits in Fig. \ref{cluster1}, the sequence of measurement and control to be implemented on the cluster needs to be determined. Another example using the cluster in Fig. \ref{cluster1} is shown in  Fig. \ref{cluster3}. 

The CZ entangling gates must be implemented by a conditional (post-selected) nondeterministic gate, if only linear optics is available. This can be done using the KLM teleportation scheme, but there is a better way using fusion gates of Browne and Rudolph (BR, \cite{BR}). Fusion operators are Kraus operators $\mathcal M$ (\cite{kok2010introduction}), they are not proper projectors but ${\mathcal M}^\dagger {\mathcal M}$ is a projector onto the even parity subspace.  Fusion operators take two qubits to one: $ \mathcal{M}=|0v\rangle\langle  00| +|1v\rangle\langle 11|$ where $v$ denotes the vacuum state that results when a photon is counted and thereby absorbed. The elements of the scheme are:
	\begin{itemize}
	\item code physical qubits as single photons in polarisation modes, $(H,V)$;
	\item use polarisation rotations to measure qubits in an arbitrary basis;
    \item instead of CZ gates between clusters use probabilistic {\bf fusion gates};
	\item replace two qubits with a single qubit while retaining all cluster-state bonds on each qubit.
	\end{itemize}

SPDC sources can produce two photons entangled in polarisation. It is then easy to create Bell pairs of the form
\begin{equation}
    |0\rangle|+\rangle+|1\rangle|-\rangle
\end{equation}
where $|\pm\rangle= |H\rangle\pm |V\rangle$. In a polarisation dual-rail qubit encoding, this is a  two-qubit `cluster state' of the form $|0\rangle|+\rangle+|1\rangle|-\rangle$. This is referred to as a Bell-pair cluster.

\begin{figure}[h!]
\centering
\includegraphics[scale=0.6]{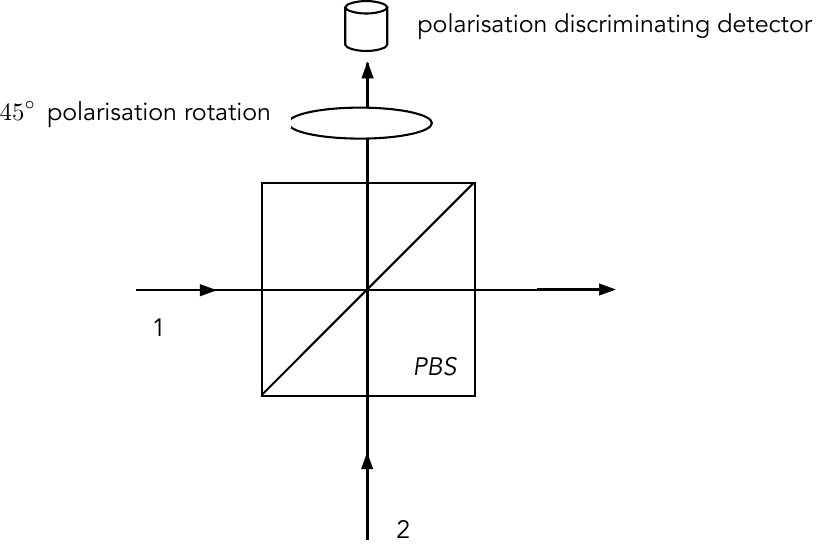}
\caption{Type-I fusion. On a polarising beamsplitter, all horizontally polarised photons are transmitted while all vertically polarised photons are reflected. This is followed by a half-waveplate that rotates $H$ and $V$ into the diagonal basis, and then a single-photon detector. This is effectively a change of  measurement basis for a polarisation-encoded qubit in the $|\pm\rangle$ basis.  }
\label{PBS-fusion}
\end{figure}

Using a polarising beamsplitter (PBS), this two-photon state may be separated into two distinct spatial and polarisation modes. This enables a Bell-pair fusion gate (Fig. \ref{PBS-fusion}). As one mode is measured,  there is only one transmitted output mode.  The state of this mode however is conditional on the photon count. 
If there is one photon (H or V) in each input of the PBS, there are three possible output counts $0,1,2$. If the input photons have the same polarisation (HH or VV), they are either both reflected or both transmitted. Only a count of $1$ is possible in either case.  This occurs with probability $0.5$.    If the photons have  different polarisation, HOM interference occurs and only  counts of $0$ or $2$ are possible.  A count of $0$ or $2$ moves out of the qubit code space for the transmitted photon, which is regarded as `failure'.  Conditioning on a single count, either $H$ or $V$, with  probability $=0.5$ implements the
conditional Krauss operation $\hat{K}_{\pm}=|0v\rangle\langle 00|\pm|1v\rangle\langle 11|$ on the original input state. Here $v$ denotes the vacuum state for the mode that is measured using an absorptive photon counter.  Failure of the fusion is heralded by the counts $(n=0,2)$. That is to say, if the input state were an arbitrary two-qubit state $|\psi\rangle_{12}$, the successful conditional states are 
\begin{eqnarray}
|\psi\rangle_{1, out}= \hat{K}_{\pm}|\psi\rangle_{12}
\end{eqnarray} 
where the vacuum state for the output of mode-2 that is subject to a photon count is omitted. 
If there are two Bell pairs, i.e. two-qubit cluster states, type-1 fusion can be used to get a three-qubit cluster state. The Bell pair
$|C_2\rangle_{12}\otimes |C_2\rangle_{34}$ has four  polarised single-photon pulses, see Fig.{\ref{Bell-pair-fuse}. As this operation is only $50\%$ successful,  four Bell pairs are used, on average,  to create a three-qubit cluster. 

\begin{figure}[h!]
\centering
\includegraphics[scale=0.5]{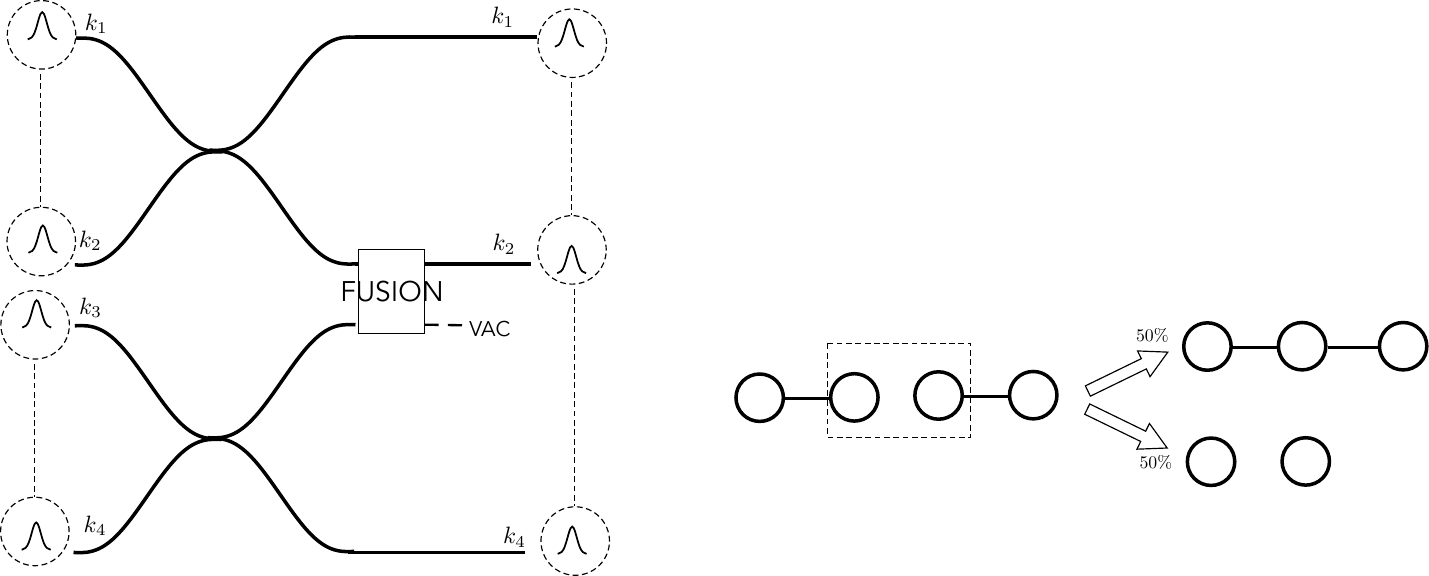}
\caption{ A scheme for fusing two  Bell-pair clusters to produce a three-node cluster. The right-hand figure gives a graphical representation of the process. When fusion fails it has the  effect of measuring both input qubits in the $ \sigma_z$ basis.   }
\label{Bell-pair-fuse}
\end{figure}

If each of the input modes encode qubits from disjoint cluster states then, when the fusion is successful, 
 the initially separate cluster qubits become a
single {\em fused} cluster qubit that inherits all the cluster
state bonds of the two qubits that were input. 
Thus, if the Type-I fusion is applied to the end qubits of
linear (i.e., one-dimensional) clusters of lengths $n$ and $m$,
success generates a linear cluster of length $n+m-1$ \cite{Psi-q}. A failure of type-I fusion effectively makes a measurement of each input qubit in the computational basis. As noted above, this  simply cuts the measured qubit from the cluster leaving all other bonds uncut. This is illustrated in  Fig. \ref{Bell-pair-fuse}.
Given a source of SPDC polarisation entangled Bell states, Type-I fusion  generates
arbitrarily long linear cluster states. 

Clearly this procedure does not increase the size of the cluster on average. However, four Bell pairs, on average,  can be used to create a three-qubit cluster. 
On average, the cluster increases by half a qubit. 
To add one qubit to the cluster requires $2\times 4-1=7$ Bell pairs, thus, $6.5$ Bell pairs per added qubit.

This enables one-dimensional cluster states to be formed.
However,  two-dimensional clusters must also be grown from linear clusters. BR (\cite{BR}) came up with an ingenious scheme for doing this using a modification of the fusion gate just discussed. The scheme is illustrated in Fig. \ref{Type-2}.
Failure of type-1 fusion (which happens 50\% of the time) is equivalent to $\sigma_z$ measurement, which breaks a bond. To get around this, the type-II fusion gate can be used. These use two Bell pairs for a resource state (i.e. a readily available source of entanglement for the computation). This is depicted in Fig. (\ref{Type-2}). Very conveniently, implementation of Type-II fusion does not require photon-number-resolving detection.
\begin{figure}[h!]
\centering
\includegraphics[scale=0.5]{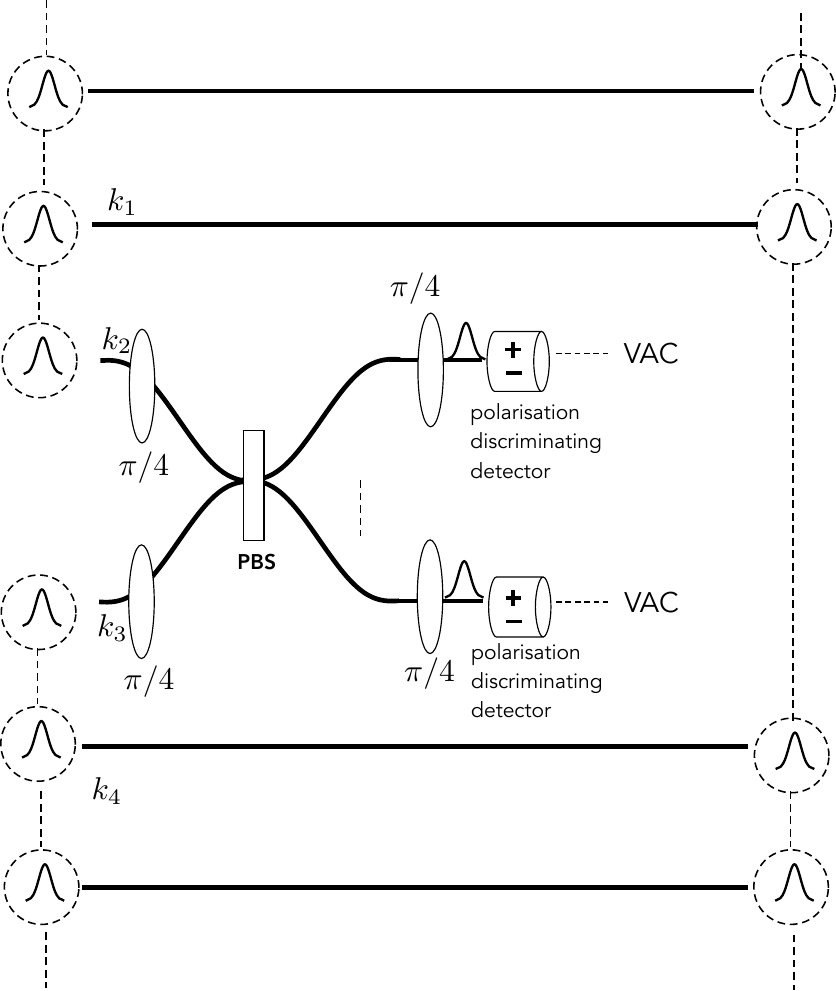}
\caption{A type-II fusion gate. Successful operation is heralded by a simultaneous  single-photon detection  event at  each photon counter. } 
\label{Type-2}
\end{figure} 
Successful operation is heralded by a single count at each photon counter. Failure is indicated if there are no photons detected at one of the counters. 
In order to understand how this can be used, BR introduced a photonic redundant encoding. This uses polarised multi-photon pulses to encode a qubit. For example,
\begin{equation}
|0\rangle_L = a_1^{\dagger}a_2^{\dagger}|0\rangle\equiv | H\rangle\otimes | H\rangle,\ \ \ \ \ \ \ |1\rangle_L = \bar{a}_1^{\dagger}\bar{a}_2^{\dagger}|0\rangle\equiv | V\rangle\otimes| V\rangle.
\end{equation}

When a Pauli $X$ measurement (i.e. measurements on the basis of $\sigma_x$, Eq. \ref{pauli} ) is made on a qubit in a cluster, it has an interesting effect. It 
does not cut the cluster, but combines the adjacent
qubits into a single two-photon encoded qubit, retaining the bonds attached to each. This can be used for redundant encoding in which a logical qubit is encoded in several qubits. The 2-qubit redundant encoding is $|{\bf 0} \rangle= |0\rangle|0\rangle$ and $|{\bf 1} \rangle= |1\rangle|1\rangle$.  A single qubit $\sigma_x$ measurement in a linear cluster joins the neighbouring qubits into a single logical qubit, redundantly encoded. 

Given the measurement scheme in Fig.\ref{Type-2}, both outputs project  onto the states  $|++\rangle+|--\rangle=|00\rangle+|11\rangle$ or $|++\rangle-|--\rangle=|00\rangle-|11\rangle$. If one of the qubits input to Type-II fusion is redundantly encoded, a fusion of the other links occurs with probability $50\%$.  Failure does not destroy the cluster but cuts it or removes qubits, requiring a further attempt.  There is no back-propagation of errors through the cluster.  By combining these two fusion gates, and using redundantly encoded multi-photon qubits, two-dimensional qubit cluster states can be generated. This leads to a dramatic cut in the number of  operations. For an arbitrary two-qubit gate, the teleportation scheme of KLM needs on average a  $100$-photon KLM teleportation resource state. Nielsen's scheme needs  54 $8$-photon KLM teleportation resource states. The Browne-Rudolph scheme  needs, on average, $52$ $2$-photon Bell states. The requirements for all LOQC schemes are: 	
\begin{itemize}
	\item high-rate source of 2-photon Bell pairs. 
	\item high-rate source of single photon.
	\item number-resolving single-photon-detectors.
	\item building optical circuits on a chip to enable scale-up. 
	\end{itemize}
A blue-print for this is presented in (\cite{Psi-q}).

\subsection{CV optical quantum computation}
\label{CV}

Encoding optical qubits in the superposition of observables with a continuous spectrum is now examined (see Section \ref{CV-encoding}). In the case of optical systems, these are the in-phase and quadrature-phase electric field operators for an optical mode defined in Eq. (\ref{canonical}).  As for photonic qubits, CV encoding is usually done using conditional (nondeterministic) gates due to the present lack of  large optical nonlinearities required to do it deterministically. In CV quantum computation, the key resource is a non-Gaussian state. 

The probability amplitude, $\psi(q)$  determines the probability density for arbitrarily accurate measurements of  the amplitude operator $\hat{q}=(a+a^\dagger)/\sqrt{2}$ as $P(q) =|\psi(q)|^2$. If $\psi(q)$ is non Gaussian, the corresponding quantum state, $|\psi\rangle$, is called non Gaussian. In such cases the Wigner representation of the state in phase space is also non Gaussian.   The corresponding probability amplitude, $\tilde{\psi}(p)$ for measurement of the canonically conjugate quadrature amplitude, $\hat{p}=-i(a-a^\dagger)/\sqrt{2}$, is given by the Fourier transform of $\psi(q)$ (see Sec. \ref{CV-encoding})and $[\hat{q},\hat{p}]=i$. In the $q$-representation, the action of these operators is (where the double arrow represents correspondence)
\begin{eqnarray}
   \hat{q}|\psi\rangle & \leftrightarrow&  q\psi(q)\\
   \hat{p}|\psi\rangle & \leftrightarrow & -i\frac{d}{dq}\psi(q)
\end{eqnarray}
This implies that the $\hat{p}$ is the generator of translations in $\hat{q}$, and $\hat{q}$ is the generator of translations in $\hat{p}$
\begin{eqnarray}
     e^{-iQ\hat{p}}|\psi\rangle & \leftrightarrow&  \psi(q+Q)   \\
     e^{-iP\hat{q}}|\psi\rangle & \leftrightarrow&  \tilde{\psi}(p+P)
\end{eqnarray}
If a state is Gaussian it remains so under these translations. More generally, the phase-space displacement operator $D(\alpha)=\exp(\alpha a^\dagger-\alpha a)$, where $\alpha=Q+iP$. This also takes Gaussian states to Gaussian states. The only other unitary operator that transforms single-mode Gaussian states to Gaussian states is the  squeezing operator $S(\xi)=\exp[(\xi ^* a^2-\xi a^{\dagger\ 2})/2] $.  In the case of two modes, entangled Gaussian states can be deterministically generated using the operators that are likewise quadratic in the two bosonic amplitude operators.  An important example is the two-mode unitary
$
    U(\theta)= \exp[-i\theta\hat{q}_1\hat{q}_2].
$
In many ways, this is the CV equivalent of a $ZZ$ gate. The quadratic operators are analogous to the beamsplitters and phase-shifters. For a thorough discussion of these unitary Gaussian transformations, see (\cite{PhysRevA.100.052301}).

Braunstein and  Lloyd (\cite{PhysRevLett.82.1784}) proved that we need to add a generator that is a polynomial in the amplitude operators, $a,a^\dagger$ with order higher than two. Cubic would suffice (\cite{hillmann2020universal}), but the simplest for optical systems is the Kerr nonlinearity $(a^\dagger a)^{2}$.  However, it can also be achieved using nonunitary (measurement-based) transformations that describe state transformations under conditional measurement.  This shifts the nonlinearity into the measurement interaction itself. 

Measurement of the canonically conjugate quadrature amplitude operators $\hat{q},\hat{p}$ is done via homodyne detection. This requires mixing the signal with a coherent local oscillator (LO) field on a balanced beamsplitter. The LO has the same carrier frequency as the signal but is phase-shifted by $\theta$. If the resulting homodyne current is integrated, the result is a real random variable corresponding to a measurement of the operator $\hat{X}(\theta)=(ae^{i\theta} +a^\dagger e^{-i\theta})/\sqrt{2}$(\cite{Wiseman_Milburn_2009}). Thus $\hat{q}=\hat{X}(0),\hat{p}=-\hat{X}(\pi/2)$.

GKP states can be prepared non-determinisitically via homodyne detection using a {\em probe} optical mode $b,b^\dagger$ coupled to the signal mode via the interaction 
\begin{equation}
\label{displace-cross-kerr}
H=\hbar\chi \alpha (a^\dagger+ a ) b^\dagger b
\end{equation}
where $a,a^\dagger$ are the annihilation and creation operators of the \emph {signal} mode and $b,b^\dagger$ are those operators for the probe mode. This interaction is thus linear in the signal mode but quadratic in the probe mode. It is the standard interaction for opto-mechanincal systems(\cite{BM}).  The signal mode is prepared in a  squeezed state
\begin{equation}
|\psi_{in}\rangle =S(r)|0\rangle
\end{equation}
and the probe mode is prepared in a coherent state $|\beta\rangle$ with $\beta \in {\mathbb C}$ (i.e. $\beta$ belongs to the set of complex numbers ${\mathbb C}$). 
A phase convention is chosen such that $\alpha $ is real and $r>0$. This implies  a phase-squeezed state, that is to say, the quadrature phase amplitude in phase with the coherent displacement is anti-squeezed.

After a fixed interaction, the probe mode is projected into an eigenstate of $\hat{q}_b=(b+b^\dagger)$. This is equivalent to homodyne detection.  If the result of this measurement is $y\in {\mathbb R}$, the (unnormalised) conditional state of the signal mode is given by
\begin{equation}
|\tilde{\psi}_{out}\rangle = \Upsilon(y)|\psi_{in}\rangle
\end{equation}
where the Krauss operator is given by 
\begin{equation}
{\Upsilon}(q|y)=(2\pi)^{-1/4}\exp\left [iyr\sin(\theta q-\phi)-(y-2r\cos(\theta q-\phi))^2/4\right ]
\end{equation}
This is a conditional probability amplitude, and 
 \begin{equation}
 \left |\Upsilon(q|y) \right |^2=(2\pi)^{-1/2}\exp\left [-(y-2r\cos(\theta q-\phi))^2\right ].
\end{equation}



 The unnormalised conditional output state, conditioned on the measurement result $y$, in the diagonal representation of $\hat{q}$ is 
 \begin{equation}
 \tilde{\psi}^{(\phi)}(q|y)=\Upsilon(q|y) \psi_{in}(q)
 \end{equation}
 Normalising the state gives
  \begin{equation}
\psi^{(\phi)}(q|y)= \frac{1}{\sqrt{{\cal P}(y)}}  \Upsilon(q|y) \psi_{in}(q)
 \end{equation}
 where
 \begin{equation}
{\cal P}(y) = \int_{-\infty }^\infty dq\   \left |\Upsilon(q|y) \right |^2\ |\psi_{in}(q)|^2.
 \end{equation}

 The $q$-amplitude probability density of the conditional output state is shown in the right-hand side of Fig. (\ref{GKPencoding}).   This shows a finite-energy approximation to a GKP state, which ideally consists of infinitely squeeze states. However, such a state requires infinite energy, hence these are approximated as shown in Fig. (\ref{GKPencoding}).

\subsection{Gaussian boson sampling}
\label{GBSsection}
A very important class of non-Gaussian states can be generated using Gaussian boson sampling (GBS) (\cite{GBS}). This relies on the ability to prepare  single-mode squeezed states, entangled by  Gaussian unitary transformations (an LOQC network), with number-resolving photon counting (i.e. a detector that can give the number of photons, as opposed to a detector that just distinguishes between zero and non-zero photon count)  for post-selection (\cite{Xanadu-BP}). This approach can produce cluster states using GKP states, enabling MBQC in a fault-tolerant way, at least in principle.

A simple scheme for producing a small GKP state using GBS is shown in Fig. \ref{GBS-GKP2}.
\begin{figure}
    \centering
    \includegraphics[scale=0.6]{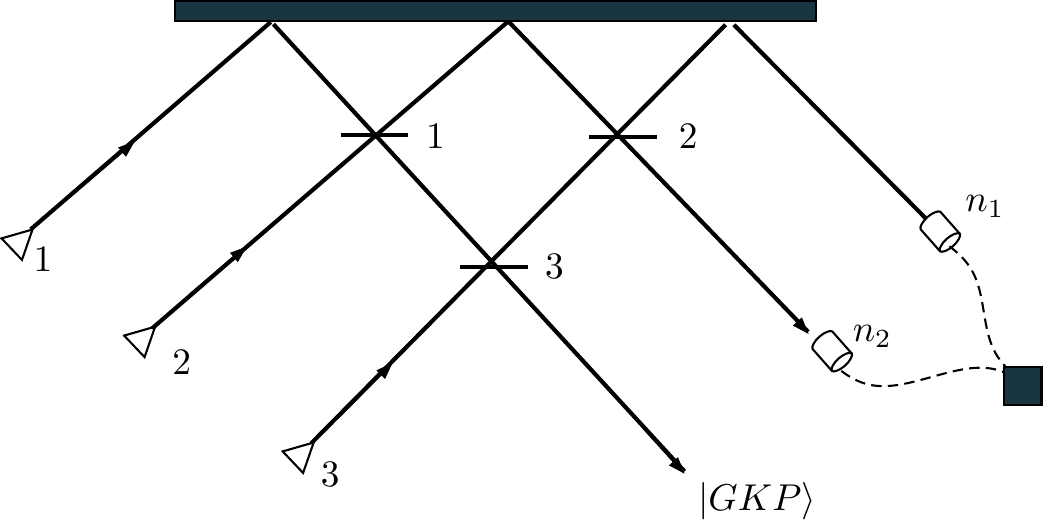}
    \caption{An optical Gaussian Boson Sampling scheme for producing an approximation to a GKP state, when the output string in the detectors is $(n_1,n_2)$. The inputs are single-mode displaced squeezed vacuum states.  }
    \label{GBS-GKP2}
\end{figure}
The analysis of this GBS scheme is given in (\cite{PhysRevA.100.052301}). The output state (which is heralded, i.e. one knows when such output state occurs) is a superposition of number states, up to a maximum total number ($n_{max}$), followed by a squeezing operation, parametrised by $\xi_1$ and a displacement, parametrised by $d_1$, 
\begin{equation}
    |\psi\rangle_{out}\rangle =D(d_1)S(\xi_1)\sum_{n=0}^{n_{max}}c_n|n\rangle.
    \label{GBS}
\end{equation}
By carefully adjusting the parameters of each of the three beamsplitters, the output state can be made a good approximation to a GKP state with a Gaussian envelope for a particular count pattern. 

An early proposal using nondeterministic teleportation gates for cat-state codes was given in (\cite{Ralph2003}, see Sec. \ref{CV-encoding}). Many more schemes have been proposed. Parity cat states are defined as
\begin{equation}
    |\pm\rangle ={\cal N}_{\pm}(|\alpha\rangle\pm|-\alpha\rangle)
    \end{equation}
    where the normalisation constant is given by ${\cal N}_{\pm}^{-1/2}=\sqrt{2} \left (1\pm e^{-2|\alpha|^2}\right )$. They are eigenstates of the parity operator $\Pi= e^{-i\pi a^\dagger a}$. The qubit code is then simply $|0\rangle=|+\rangle, |1\rangle=|-\rangle$. 
Cat states produced nondeterministically can easily be entangled on beamsplitters (\cite{Ralph2003}), thus entangling gates are easy. Single qubit gates are more difficult but can be done using weak displacements and teleportation (\cite{Ralph2003}).   Kerr nonlinearities can produce cat states with Poisson statistics (\cite{Mil-Fredkin}) which can be large in superconducting quantum circuits, but are too small for optical systems. In that case nondeterminsistic (measurement-based) protocols can be used(\cite{PhysRevA.103.013710}). These only require single-mode squeezed states and number-resolving photon counting. This approach is promising for scalability(\cite{Xanadu-BP}).

\subsection{Detection}

Detection in photonic quantum computing can be divided into homodyne detection and photon-number detection. In both cases, a photodetector converts photons to electrons, and the resulting electric current is processed to give either a quadrature measurement  or a photon number.

As mentioned in Sec. \ref{CV}, the quadrature amplitude operators $\hat{q}$ and $\hat{p}$ are measured via homodyne detection. The signal to be measured is spatially overlapped on a 50/50 beamsplitter with a local oscillator field  (the probe mode) that is of the same frequency but phase-shifted by $\theta$ (Fig. \ref{homodyne}).  In practice, the LO field could come from the laser that pumps the nonlinear crystal that generates the squeezed states. The LO can be phase-shifted simply by adjusting the length of one arm of the interferometer or by fine control of a glass plate in the LO arm. The expectation value of the photocurrent (normalised by the amplitude of the LO)  $J_{hom}(t)$ is

\begin{eqnarray}
      J_{\text {hom }}(t)& = & \langle e^{i\theta} a(t)+e^{-i\theta} a(t)^\dagger \rangle +\xi(t) \\
      & = &  \sqrt{2} \langle\hat{X}_\theta(t)\rangle+\xi(t) 
      \label{current}
\end{eqnarray}
where $\hat{X}_\theta(t) =\hat{q}(t)\cos\theta -\hat{p}(t)\sin\theta $
and $\xi(t)$ is white noise. The photocurrent is proportional to the expectation value of the quadrature $\hat{q}$ (\cite{Wiseman_Milburn_2009}).
\begin{figure}
\centering
\includegraphics[scale=0.35]{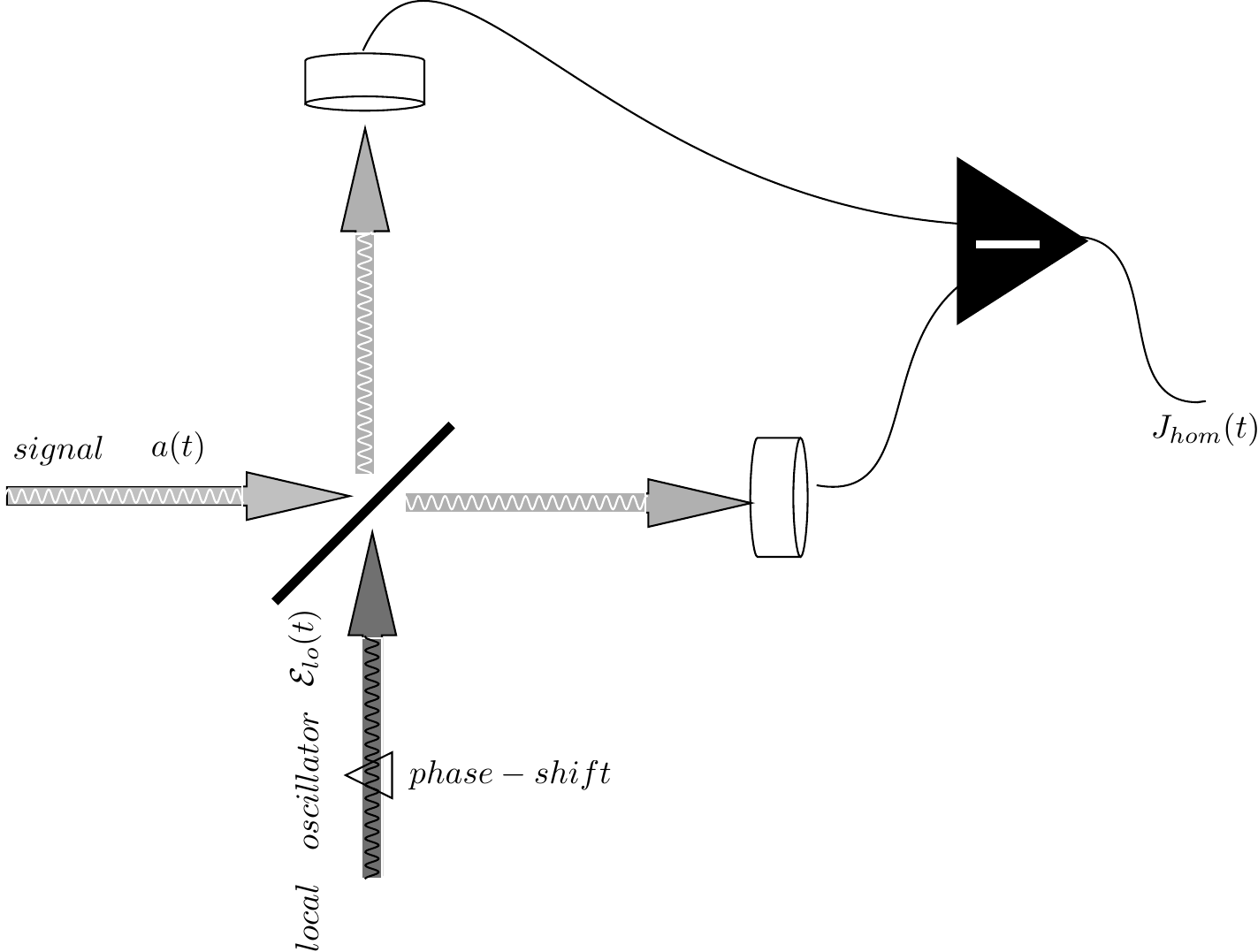}
\caption{Balanced homodyne detection. A weak signal field with positive frequency components $a(t)$ is combined with a strong local oscillator in a coherent state with the same carrier frequency as the signal field. Both outputs are directed to photodetectors and the resulting currents combined by subtraction in an electronic mixer. The output current is called the homodyne current.  }
\label{homodyne}
\end{figure} 
The Fourier transform of a two-time (stationary) correlation function for the homodyne current  enables a noise power spectrum $S(\omega)$ to be defined, 
\begin{equation}
  \langle :\ \Delta \hat{X}_\theta^2\ :\rangle =\frac{1}{2}\int_{-\infty}^\infty d\omega (S(\omega)-1)
\end{equation}
where the left hand side is the  normally ordered variance (\cite{Wiseman_Milburn_2009}) in $\hat{X}_\theta$.

 The integrated homodyne current,
\begin{equation}
\int_0^\infty J_{hom}(t) dt 
\end{equation}
is a random variable(\cite{Wiseman_Milburn_2009}). It can be  shown that the statistics of this random variable is determined by the marginal quantum statistics given by $|\psi_{in}(q)|^2$ (see Sec. \ref{multimode-states}).

Importantly, Eq. \ref{current} allows the measurement of either the amplitude quadrature (when $\theta{=}0$) or the phase quadrature (when $\theta{=}\pi/2$). For coherent states, which have the same uncertainty in both quadratures, there is no particular advantage to using a homodyne detector.  For squeezed states, which have different uncertainties in different quadratures,  the homodyne detector is necessary. This technique works for both bright squeezed light or squeezed vacuum states. 


Measurements in DV optical quantum computation require the detection of photons. This can be further divided into two types. The first type distinguishes between ``nothing" and ``some photons", this is also called a bucket detector. The second type can distinguish between different photon number, hence the name photon-number-resolving detector. A photon detector has an efficiency $\eta \in[0,1]$, defined as the probability that a photon count registers when a single photon is incident on the detector.  Single photon detectors are modeled as having a beamsplitter before them with a transmission coefficient of $\eta$, and reflection coefficient of $1-\eta$. The incident quantum state to a photon detector is a superposition of different number states.  The measurement describing a bucket detector, where 0 denotes no count, and 1 denotes a count, is given by the corresponding positive operator-valued measures (\cite{kok2010introduction}),
\begin{equation}
    \hat{E}_0=\sum_{n=0}^{\infty}(1-\eta)^n|n\rangle\langle n| \quad \quad \hat{E}_1=\sum_{n=0}^{\infty}\left[1-(1-\eta)^n\right]|n\rangle\langle n|.
\end{equation}
Then $ p_x={\rm tr }[E_x\rho]$ (where $x \in {0,1}$) is the distribution of the binary outcomes.

In general, a photon number-resolving detector is described by the positive operator-valued measures
(\cite{scully1969quantum})
\begin{equation}
\hat{E}_n=\sum_{k=n}^{\infty}\left(\begin{array}{l}
k \\
n
\end{array}\right) \eta^n(1-\eta)^{k-n}|k\rangle\langle k|.
\end{equation}
where $\eta$ is the conditional probability of counting a single photon given one photon is present in the mode. 

There is also a small probability that a count is registered even when no photons are incident, this leads to dark counts. A good photon detector has high efficiency and low dark counts. The most convenient single-photon detectors are silicon avalanche photodiodes (APDs) that are operated in Geiger mode. A photon hitting the APD triggers the emission of an electron into the conductance band. The electron is accelerated in an electric potential to induce an avalanche of secondary electrons. The resulting current indicates the presence of the photon. The avalanche is stopped by reversing the electric potential leading to some dead time in the detector. Typical APDs have a dead time of a few nanoseconds and dark counts that range from $\sim 10$ to $\sim 100$ a second. Silicon APDs are convenient because they operate at room temperature, but their efficiency is limited to about $\sim 65 \%$ in the mid-infrared and they have low efficiencies for telecom wavelengths.  Indium-Galllium-Arsenide (InGaAs) detectors are used for telecom wavelengths, with typical efficiencies of $10\%$ to $30\%$, and dark counts typically at $\sim 1000$s per second.

The efficiency of single-photon detection improved greatly with the advent of superconducting nanowire single-photon detectors, which consist of a meandering superconducting material on a substrate, as shown in (see Fig. \ref{SNSPD}). Superconducting nanowire single-photon detectors (SNSPDs) have efficiencies of over $\sim 90\%$ (\cite{marsili2013detecting}). SNSPDs operate near the critical current ($I_B$) to be at the superconducting phase. The absorption of a photon causes the device to have a higher resistivity, leading to a voltage spike that is considered as the photon detection (Fig. \ref{SNSPD}.c). The dead time of an SNSPD is similar to that of an APD; the timing jitter is very small (in the few ps range). The increase in efficiency comes at the price of a more complicated setup that involves a cryostat, because the operating temperatures to reach superconductivity are between 0.8 to 3 K. The specific wavelength range that the SNSPD is sensitive to is determined by the choice of the superconducting material and the optical cavity around it. SNSPDs are bucket detectors, photon-number resolution being limited by the saturation of the avalanche process. However, it has been shown recently that analysing the timing of the rising and falling edges of the output pulses give some degree of photon-number-resolution, although only for low photon numbers  ($\sim 4$)(\cite{sauer2023resolving}). 

\begin{figure}
\centering
\includegraphics[scale=1.3]{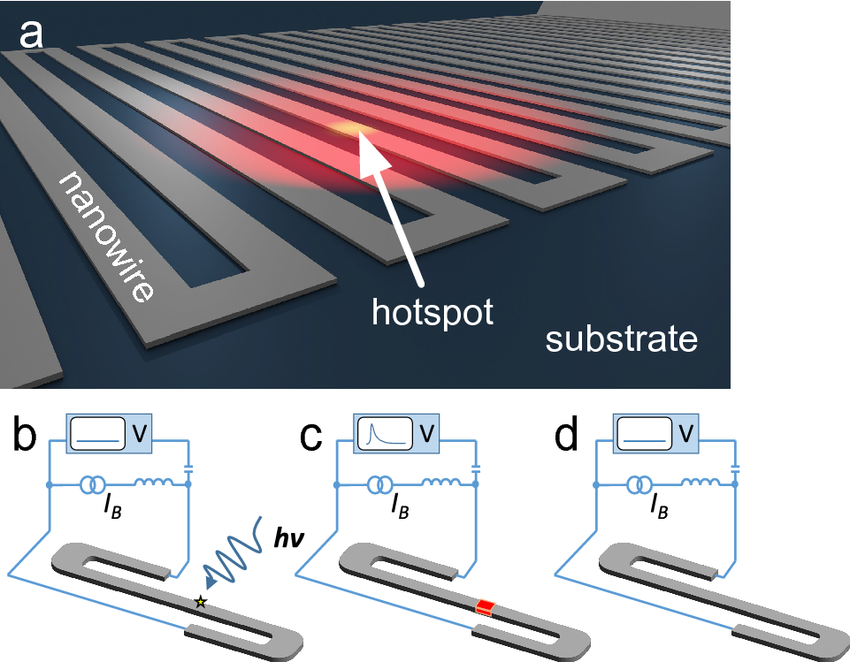}
\caption{Superconducting nanowire single-photon detectors (SNSPDs) working principle. The SNSPD is operated near the critical temperature. Absorption of a photon causes a hotspot to develop (a). The nanowire is normally superconducting and DC bias $I_B$ is applied (b). The hotspot induces additional current and the critical current density is surpassed, a voltage peak is registered (c). The heat is then dissipated, and the SNSPD is ready for detection again (d). (Figure lifted from (\cite{fuchs2022photon}), used under \href{https://creativecommons.org/licenses/by/4.0/}{CC by 4.0}.)}
\label{SNSPD}
\end{figure} 

Photon-number resolution is possible using another kind of superconducting detector---transition-edge sensors (TESs). Photon-number resolution is possible in a TES because the change in resistivity is proportional to the number of photons absorbed. The TES acts as a very sensitive bolometer. The number of photons can be read out by integrating the current produced as a result of the change in resistivity.  Recent results have shown photon-number resolution of up to $\sim 20$ (\cite{harder2016single}). The efficiency of a TES is comparable to that of SNSPDs. The dark-count rates are lower, but the dead time (in the $\mu$s range) and timing jitter (in the ns range) are greater than that of an SNSPD (\cite{lamas2013nanosecond}). As seen in Sec. \ref{GBSsection}, photon-number-resolution is needed in the preparation of the GKP states, hence the great interest in developing this type of detectors.

One subtlety that deserves mentioning is that the detection for CV optical quantum computing assumes that the input and output modes are perfectly mode-matched---the detector is sensitive to all the input modes.  This is different from the case of DV where some mode filtering (e.g. a polarising beamsplitter or a multiplexer) that depends on the encoding used is often placed before the photodetector.

\section{Large-scale Photonic Quantum Computing}
\subsection{Single photon sources}
\label{single-photons}
 Scalable optical quantum computing requires single-photon sources. As discussed in Sec. (\ref{multimode-states}),  a single-photon state is a coherent superposition of a single excitation over many frequency modes(\cite{Raymer-Walmsley}). The key signature of a true pure-state single-photon source is provided by the Hong-Ou-Mandel (HOM) interference (\cite{HOM}) discussed in Sec. \ref{prelim}.  
                                             
A good example of a single-photon source that exhibits both a suppression of second-order correlation at zero time and good HOM interference visibility was demonstrated by the group of Rempe(\cite{Rempe}) in Garching and also by the group of Kuhn at Oxford (\cite{Kuhn2011}).  These experiments verified the temporal shape of their photon amplitudes and phases in detection-time resolved HOM and quantum homodyne experiments.

For scalable optical quantum computing, single-photon sources need to be integrated into a device.  Single-photon sources for measurement-based protocols, such as fusion,  are being developed across a number of technologies,  such as doped fibres(\cite{Shimizu}), diamonds(\cite{diamond}), and semiconductor-based sources(\cite{Senellart}). The performance benchmarks require very good control over rate of emission and control of the spatio-temporal mode functions of each photon. These translate into an efficiency metric for achieving nearly pure quantum states over many trials.  There are already many commercial single-photon source developers, e.g. the quantum-dot photon source in (\cite{maring2024versatile}) has an indistinguishability of $\geq$99.5\% (i.e. the photons generated are identical in their properties, such that one photon cannot be distinguished from another).  An architecture based on single-photon sources is presented in (\cite{maring2024versatile}), with a $>$99\% single-photon- and $>$93\% two-photon-gate fidelity.   New technologies for quantum optical sources are likely to emerge in coming years.

\subsection{Error correction}
 For all quantum-computing technologies, the challenge is to implement error correction to enable large-scale fault-tolerant computation(\cite{brun2020quantum}).  The discovery of quantum error correction by Calderbank and Shor (\cite{Shor}) and independently by Steane (\cite{Steane}) was the main reason why quantum computing has not remained little more than a curious quantum phenomenon and instead has moved to the technological frontier attracting billions of dollars in funding. The largest technological challenge for optical QC schemes is photon loss. The solution is to understand photon loss as a qubit error and find error correction codes to reduce the probability for that error (\cite{brun2020quantum}).

 Much is known about the kinds of code that will be useful in quantum optical schemes. The fusion gate approach suggests a number of new ways to implement surface codes in measurement-based quantum computation. The error-correction methods developed by PsiQuantum can tolerate a 2.7\% probability of suffering a single-photon loss(\cite{Psi-q}). These methods,  implemented in integrated photonic devices, offer a clear path to very large-scale QC. Measurement-based quantum computing for continuous variables under development at {\it Xanadu}   also suggest specific quantum error correction protocols. The scale-up of GBS machines is proceeding rapidly(\cite{Madsen}). These offer significant non-classical capability without error correction(\cite{Walmsley:23}).

\section{Conclusion and Outlook for Photonic Quantum Computing}

There is a fundamental difference between optical qubits and matter qubits. Matter qubits are usually encoded at fixed spatial locations (except for some ion-trap schemes). Optical qubits are encoded in time. This difference parallels the division of classical information processing into computation and communication. Optical quantum computing shows that communication with quantum  sources and detectors is just as powerful as a quantum computer that uses  qubits encoded in the physical states of matter.   

The word `qubit', like `information', is an abstract noun. It does not refer to a physical thing. It refers to a process by which information is encoded in physical systems. When seeking to encode a qubit in the physical world the encoding cannot be separated from the decoding; there is no point in choosing a particular physical basis to encode a qubit if no sensible measurement scheme exists that can make efficient measurements in that basis. 

 In 2000, Di Vincenzo listed a set of criteria that is required for the realisation of quantum computation (\cite{divincenzo2000physical}). These criteria are rooted in the circuit model of quantum computation but not limited to the circuit model. Having laid out the basics of photonic quantum computing have been laid out, the following paragraphs list these criteria and evaluate how a photonic quantum system fares against them:
\\
\noindent\emph{\textbf{Well-defined state space of qubits.}} Quantum states of light enable qubits to be encoded, as discussed in Sec. \ref{optical-qubits}. In optics, the state space is the tensor product over the Fock states of each orthogonal mode. This should be contrasted with matter qubits, where the state space is the tensor product of the Hilbert space for each particle.    Taking polarisation as an example of a qubit encoding, an off-the-shelf polarizing beamsplitter distinguishes between the horizontal and vertical polarisations with an error in the order of $10^{-5}$.   This is a very low measurement error for a binary process; such measurements are called `efficient'.  However, to do this at the single-photon level is more difficult, as it requires number-discriminating single-photon detection.  The technology to do this is now well developed so it makes sense to encode a dual-rail photonic qubit in two modes distinguished by orthogonal polarisation.


\noindent\emph{\textbf{Ability to initialise the state of the qubits.}} The system of qubits that will be used for computation should be initialised to a known quantum state, just in the same way that a classical register is initialised to a known state. In optics this requires preparing the state of each mode in a known pure state. For dual-rail photonic encoding this is not easy but the technology of single-photon sources is now well developed. The challenge is entangling states of different modes. As has been seen, non-deterministic but heralded processes can achieve this.    Recent developments in deterministic single-photon generation and photonic quantum gates inspire some optimism.

\noindent\emph{\textbf{Long relevant decoherence times.}} A quantum system in contact with an environment will eventually decohere. In the most extreme case, the decoherence time may be taken as the time it takes for a superposition, $c_0\ket{0}+c_1\ket{1}$, to become practically indistinguishable from the mixture, $|c_0|^2\ket{0}\bra{0}+|c_1|^2\ket{1}\bra{1}$. In the case of a single-photon pulse, a transform-limited pulse corresponds to a pure state and decoherence corresponds to additional phase noise. This is very low in typical optical fibres. Transform-limited pulses remain transform-limited for very long propagation times. Unfortunately photons are easily lost by absorption or scattering. This is the real source of decoherence and error in photonic quantum computing. Fortunately photon loss is often heralded using number-discriminating detectors.


\noindent\emph{\textbf{A universal set of quantum gates.}} Quantum computation is typically a sequence of unitary transformations acting on one or more qubits. A set of quantum gates is universal if composing gates belonging to the set enables any unitary transformation to be approximated with arbitrary accuracy. The specification of universal gate sets is agnostic to the physical system used for quantum computation but the implementations are highly dependent on the physical systems. In the case of matter-based qubits the physical systems are fixed in place and time-dependent Hamiltonians are switched on and off in a particular order corresponding to the algorithm being implemented. In the case of photonic qubits, the gates are stationary  and  the qubits are travelling pulses. The gates are implemented by synchronising pulse `collisions' on various optical elements such as beamsplitters.

As discussed in Sec. \ref{gates}, any arbitrary single-qubit gate can be represented by one- and two-mode optical transformations as combinations of beamsplitters and phase-shifters—an optical interferometer. The time to do a sequence of single-qubit operations is short, simply the time it takes for the photon to traverse the interferometer.  Multi-qubit gates are more of a problem.  The fact that optical photons do not interact is a great challenge; it means that it is difficult for the state of one mode to be directly coupled to the state of another mode.  Using the optical Kerr nonlinearity is a way to do photonic two-qubit gates, but this is weak and lossy.  The most promising approach so far is based on linear optics and heralding measurements. The quality of a gate is then  determined by the efficiency of the measurement.   The success rate is not 100\%, but the success of the two-qubit gate is heralded.

\noindent\emph{\textbf{A qubit-specific measurement capability}}.  The result of the computation can be obtained from the final state of the quantum system.  The final state needs to be measured in order to obtain the classical information, e.g. “0” with probability $p$, and “1” with probability $1-p$. Projective measurements are unphysical; it is efficient measurements that are important. An efficient measurement is a highly accurate measurement and is necessary for quantum computation.  Efficient measurements of photonic quantum states are relatively well-developed for a variety of photonic properties because of the availability of homodyne detection, mode-filtering, and single-photon detectors that are of very high efficiency.  Photodetectors used for homodyne detectors have over 99\% efficiency at wavelengths around 1550 nm. The highest overall detection efficiency achieved with bulk optics so far is 97.5\%, at 1064 nm (\cite{vahlbruch2016detection}).  Superconducting nanowire single-photon detectors have over 90\% efficiency, with (\cite{alexander2024manufacturable}) reporting a 93\%  median on-chip efficiencey at 1550 nm.


Photonic quantum computing is on a path to controlling hundreds of error-protected logical qubits (\cite{Xanadu-BP},\cite{alexander2024manufacturable})---certainly a challenging path but a path nonethless. No matter what technology platform eventually wins in the market for commercial quantum computing, optical technology will always be required for quantum interconnects.  It will be advantageous to do some computational processing  in the optical in-line networks. In the early 1980s, much was made of the possibility of using classical optics for computation. It did not go away; it became incorporated into the optical-fibre communication system. Photonic quantum computation may be similar (\cite{mcm}).

\section{Alternative Computing Architectures and Further Reading}

There is a developing awareness that the current paradigm for quantum computation based only on  qubit circuits  may not be the best approach to harnessing the quantum world for computational tasks, especially machine learning(\cite{Fernandez}).  The current quantum computing paradigm closely follows the classical computational model based on CMOS digital logic circuits and von Neumann architecture. The rise of machine learning provides another impetus to reconsider the current computational paradigm. The cost of machine learning is largely associated with the energy cost of implementing a von Neumann architecture on CMOS digital computers(\cite{Chojnnacka}).

In recent years the idea of using physical systems, especially quantum systems, to perform computational tasks has resurfaced(\cite{Datta,coles2023thermodynamic, McMahon}). Of particular interest is the approach of {\it Normal Computing} and {\it Extropic} (\cite{ coles2023thermodynamic, extr}). They give a framework for physics-inspired AI algorithms  based on probabilistic switching in physical devices. Machine-learning algorithms can be viewed as data analysis and statistical tools for finding patterns in data. There are quantum-optical versions of deep neural networks(\cite{optical-DNN, info15020095,ma2023quantumnoiselimited}), convolutional neural networks(\cite{zapletal2023errortolerant}) and kernel embedding (\cite{Bowie_2024}).   

A very early physical machine approach was developed by DWave in the early 2000s. It is best described as an annealing computational machine using dissipative quantum tunnelling between localised states(\cite{Dwave}). It is based on finding steady states that are close to many-body ground states that happen to solve an optimization problem. However, there is no reason to restrict quantum steady states to such states. Far-from-thermal-equilibrium steady states offer greater possibilities and quantum phase transitions at low temperature can be easily exploited in quantum optics.  An example of this is the coherent Ising machine approach developed by Yamamoto and coworkers (\cite{CIM}), but has yet to reach a clear quantum advantage. 

An exciting recent development is the rise of hardware implementations of machine learning using optical systems. These will move rapidly to quantum-optical demonstrations.  Raman quantum memories, in atomic(\cite{atomic-memories}) or optomechanical systems(\cite{Lake}), offer a possible path to quantum-convolutional neural networks, if they can be made to work at the single-photon/phonon level. The technological dividend of single-photon computing could deliver a huge increase in the thermodynamic efficiency of AI (\cite{ma2023quantumnoiselimited,McMahon}).  

\bibliography{ORE_final}
\end{document}